\documentclass[fleqn,usenatbib]{mnras}

\DeclareRobustCommand{\VAN}[3]{#2}
\let\VANthebibliography\thebibliography
\def\thebibliography{\DeclareRobustCommand{\VAN}[3]{##3}\VANthebibliography}

\usepackage[normalem]{ulem}
\usepackage{gensymb}

\usepackage{graphicx}	
\usepackage{amsmath}	
\usepackage{amssymb}	
\usepackage{lmodern}
\usepackage{newtxtext,newtxmath}
\usepackage[T1]{fontenc}

\title[Mrk 876: X--Ray Broadband Spectroscopy ]{Hard X-Ray broadband spectroscopy of Mrk~876: characterizing its spectrum}

\author[Eugenio Bottacini]{
Eugenio Bottacini,$^{1,2,3}$\thanks{E-mail: ilbotta4@gmail.com}
\\
$^{1}$W.W. Hansen Experimental Physics Laboratory \& Kavli Institute for Particle Astrophysics and Cosmology, Stanford University, USA\\
$^{2}$Dipartimento di Fisica e Astronomia "G. Galilei", Universit\`a di Padova, Padova, Italy\\
$^{3}$Eureka Scientific, 2452 Delmer Street Suite 100, Oakland, CA 94602-3017, USA
}

\date{Accepted 2022 June 25. Received 2022 June 25; in original form 2021 December 2}

\pubyear{2022}

\begin{document}
\label{firstpage}
\pagerange{\pageref{firstpage}--\pageref{lastpage}}
\maketitle

\begin{abstract}
Ever since the launch of the {\em NuSTAR} mission, the hard X-ray range is being covered to an unprecedented
sensitivity. This range encodes the reflection features arising from active galactic nuclei (AGN). Especially,
the reflection of the primary radiation off the accretion disk carries the features of the manifestation of
General Relativity described by the Kerr metric due to rotating supermassive black holes (SMBHs). We show
the results of the broadband analyses of Mrk~876.
The spectra exhibit the signature of a Compton hump at energies above 10 keV and a broadened and
skewed excess at energies $\sim$6 keV. We establish this spectral excess to be statistically significant at
99.71\% ($\sim$3$\sigma$) that is the post-trail probability through Monte Carlo simulations. 
Based on the spectral fit results and the significance of spectral features the relativistic reflection model
is favored over the distant reflection scenario. The excess at $\sim$6 keV has a complex
shape that we try to recover along with the Compton hump through a self-consistent X-ray reflection model.
This allows inferring an upper limit to the black hole spin of $a$~$\leq$~0.85, while the inclination angle of the
accretion disk results in $i$=32.84$\degree{}^{+12.22}_{-8.99}$, which is in agreement within the errors with
a previous independent measurement ($i$=15.4$\degree{}^{+12.1}_{-6.8}$).
While most spin measurements are biased toward high spin values, the black hole
mass of Mrk~876 
(2.4$\times$10$^{8}$~M$_\odot$~$\leq$~M$_{SMBH}$~$\leq$~1.3$\times$10$^{9}$~M$_\odot$)
lies in a range where moderately spinning SMBHs are expected.\\
Moreover, the analyses of twelve {\em Chandra} observations reveal for the first time X-ray variability of Mrk~876 with
an amplitude of 40\%.
\end{abstract}


\section{Introduction}
Several precise measurements of the cosmic X-ray background \citep[CXB; e.g.][]{faucher20} confirm a common spectral feature: its peak at
$\sim$30 keV. Astronomers and astrophysicists agree that most, if not all, of the CXB radiation at its peak is the result of the integrated emission
by active galactic nuclei (AGN). Their accretion efficiency, and ultimately their luminosity, strongly depend on the black-hole spin: the more rapidly the black hole
spins, the more efficiently it accretes resulting in the spectral Compton hump above 10 keV \citep{george91}. Its emission peak coincides well with the
peak of the CXB. In fact, in a detailed study by \cite{vasudevan16} the authors find that 50\% of the CXB can be produced by a mere
15\% of AGN having a close to maximally spinning supermassive black hole (SMBH). The role of moderately spinning SMBHs remains an open
question as half of the spin measurements in AGNs return values close to maximally spinning black holes \citep[spin $a$~$>$~0.9;][]{vasudevan16}.
More recent and updated reviews on this subject regarding observational constraints of spin measurements from X-ray data are available in literature
\citep[e.g.][]{reynolds19, bambi21a, bambi21b, reynolds21}.
However, this contribution to the CXB might come at the expense of the number of the elusive Compton-thick AGN that have a high column 
density of the 
obscuring torus equal or larger to the inverse of the Thomson cross-section
(N$_{H}$~$\geq$~$\sigma$$_{T}^{-1}$~$\simeq$~1.5$\times$10$^{24}$~cm$^{-2}$). Such a column density is able to block a large fraction of the radiation
from the central region, which in turn hampers the detection of these sources. However, their integrated radiation is predicted to contribute at different
percentages in different population synthesis models \citep[e.g.][]{gilli07, treister09, ananna19}. Nevertheless, the presence of a large number of
Compton-thick AGN could be reconciled with the intensities of both, the CXB and the IR background \cite{comastri15}. 
Observational studies with {\em NuSTAR} show that the cut-off energy of the continuum characterizes the non-thermal spectrum. Such features are
constrained to be well above the Compton hump and even outside the range covered by {\em NuSTAR} as shown in individual source studies and
for samples of AGN \citep{fabian15, fabian17, tortosa18, middei19, balokovic20} reinforcing the importance of the Compton hump.
Given the reasons above, every case study of broadband measurements to characterize the AGN emission mechanism is of fundamental importance.\\
Regarding the emission mechanism, the primary radiation is emitted by the hot corona surrounding the SMBH.
This is seen as a power-law spectrum by an observer and possibly subject to absorption by the torus depending on the geometric arrangement. This
primary radiation also irradiates the accretion disk that reflects it through several reprocessing steps \citep[e.g.][]{matt91} including the fluorescence
emission of K$\alpha$ of most abundant elements \citep{george91}. Most emission lines are unresolved at low X-ray energies resulting in the so-called
soft X-ray excess below $\sim$1 keV. Among emission lines from abundant elements in AGN spectra, the most prominent is the Fe--K$\alpha$ line at
6.4 keV. The line can be subject to Doppler shifts, relativistic beaming, and gravitational redshift that can cause distorted line profiles and photon
energy shifts \citep{fabian00}, which allow for inferring the spin of the SMBH.
Additionally, also the electron scattered primary radiation off the inner accretion disk predicted in forma of a broad Compton
hump at energies above 10 keV \citep{ross93} can now be measured due to the superb, for this energy band, sensitivity of {\em NuSTAR} 
observations as in \cite{risaliti13}, whose authors inferred a rapidly rotating black hole. A
moderately rotating black hole ($a$~=~0.5) in Swift J2127.4+5654 was discovered by \cite{marinucci14} with very deep {\em NuSTAR} observations of
340 ksec. Such moderately spinning black holes are important because they represent the missing population in systematic spin measurement 
studies \citep{reynolds16}. In fact, many sources have lower limits compatible with much higher spin \citep{vasudevan16}.
Lower spins have not been measured yet, even though low and moderate spins need to be fully accounted for in any population statistics from spin data
\citep{reynolds16} making such spin measurements indispensable pieces in putting the puzzle together. The mass of the SMBH in Mrk~876 has been
independently constrained to be between
M$_{SMBH}$=2.4$\times$10$^{8}$~M$_\odot$) \citep{kaspi00} and M$_{SMBH}$=1.3$\times$10$^{9}$~M$_\odot$)
\citep{bian02}. Such a high mass value places Mrk~876 in an interesting parameter space of the mass--spin plane \citep{vasudevan16}, where the
authors predict intermediate spins can be found.\\
Mrk~876 is an optically selected Seyfert type-I AGN from the Palomar-Green (PG) Bright Quasar survey \citep{schmidt83} being often also referred
to as PG 1613+658. The source is interacting with another spiral galaxy \citep{yee87, hutchings92}. At X-ray energies the source has been
detected by {\em Ginga} \citep{lawson97} and later also at even higher energies in the combined {\em Swift} -- {\em INTEGRAL} X-ray (SIX)
survey \citep{bottacini12}. Successively also the survey of {\em Swift} alone \citep{baumgartner13} has detected the source, while Mrk~876 is not detected in the survey of
{\em INTEGRAL}. {\em Swift}/XRT follow-up observations suggest the source hosts a spinning SMBH measured through a transient gravitationally
redshifted Fe line \citep{bottacini15} as measured also in other AGN spectra \citep[e.g.][]{nardini16}. The very same analyses of the {\em Swift}/XRT
observations and analyses of observations by {\em XMM-Newton} \citep{porquet04, piconcelli05} reveal the absence of any absorption in excess
to the Galactic value.\\
In this research we present the analyses of the {\em NuSTAR} observation on Oct. 22nd, 2020 of Mrk~876
with the aim of characterizing its emission mechanism through its broadband spectrum. The redshift of Mrk~876
$z = 0.1385$ \citep{lavaux11} corresponds to 551.4 Mpc for $H_{0} = 73$ km s$^{-1}$ Mpc$^{-1}$ assuming Hubble flow that we do throughout
this paper.

\section{{\em NuSTAR} Observation}
\subsection {Data Analysis}
\begin{figure}
\includegraphics[width=1.0\columnwidth]{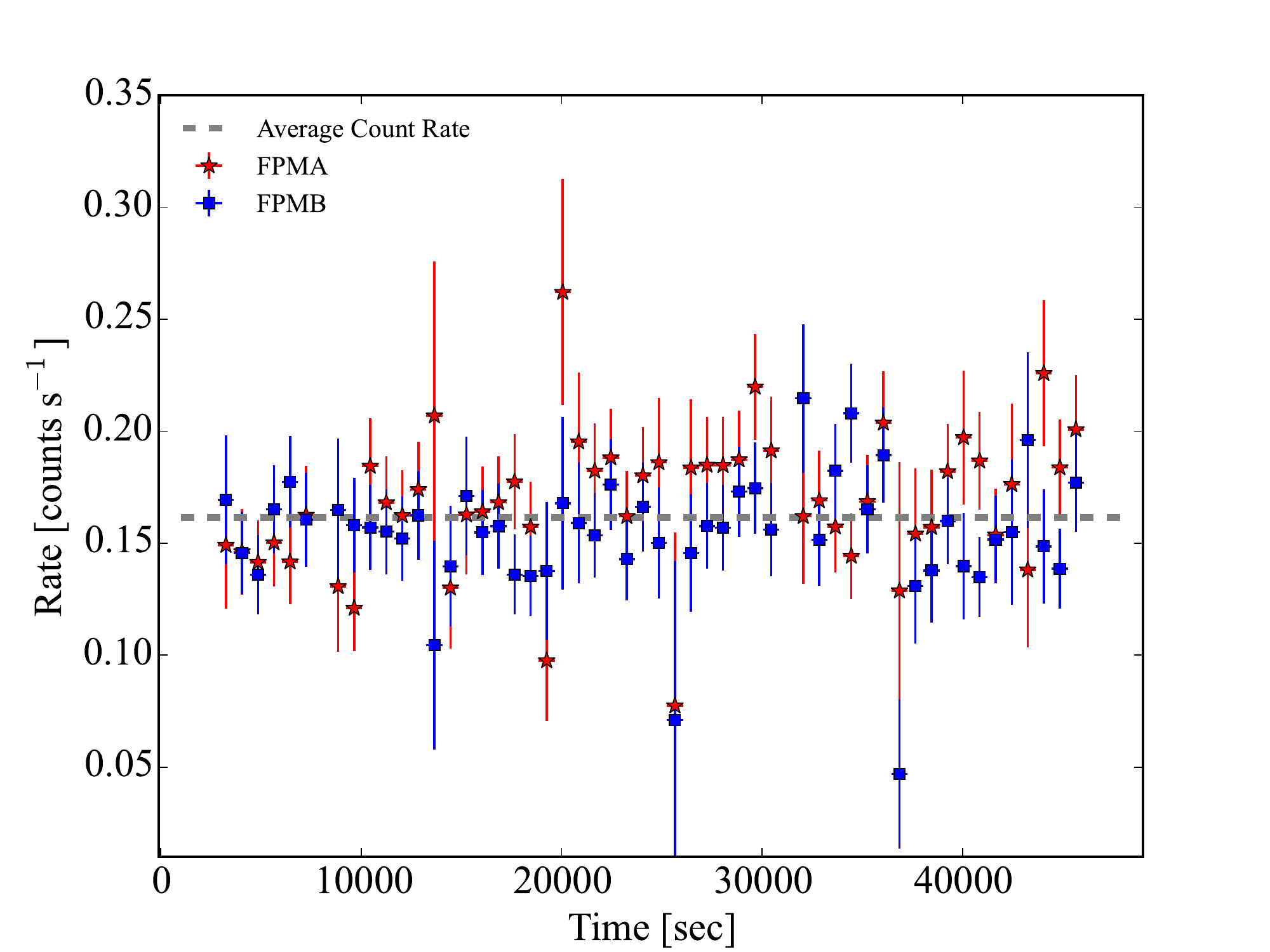}
\caption{Mrk~876's light curve binned to 800 seconds for FPMA (red) and FPMB (blue).}
\label{fig:lc}
\end{figure}
The {\em Nuclear Spectroscope Telescope Array} \citep[{\em NuSTAR};][]{harrison13} mission has two telescopes that are co-aligned and that
have two independent focal plane modules A and B (FPMA and FPMB). The  telescopes are able to focus, for the first time, hard X-rays in the
energy range 3 -- 79 keV allowing for precise imaging and sensitive spectral analyses in this non-thermal energy range.\\
Since the early days of the mission, Mrk 876 was a target of {\em NuSTAR}'s extragalactic survey program EGS, which was later
updated with target observations. This observation was performed on 2020-10-22 UT 05:46:09 for $\sim$30 ks with observation id
60160633002. The observation has been analyzed using HEASoft 6.28. Events have been reprocessed with the latest {\em NuSTAR}
Data Analysis Software (NuSTARDAS v. 1.9.6) using {\texttt{nupipeline}} and making use of its latest calibration data base (CALDB)
files version 20210427. None of the resulting two images are affected by stray-light effect \footnote[1]{https://heasarc.gsfc.nasa.gov/docs/nustar/analysis/nustar\_swguide.pdf}.
For the spectral analyses the circular extraction region extends for 60$^{\prime\prime}$ around the source position, while to maximize the
collected background data the area extends for 100$^{\prime\prime}$ avoiding any overlap with the source.

\subsection{Timing Analysis}
At hard X-ray energies ($>$ 15 keV) the {\em Swift} 105-Month Hard X-ray Survey \citep{oh18} reveals a steady light curve of Mrk~876. 
By using {\em Swift}/XRT and {\em XMM-Newton} at lower X-ray energies ($<$ 10 keV) this source has shown some flux variability
(factor of $\sim$1.6 at most) on long time scales between 1991 and 2013, while on shorter time scales of weeks, days, and hours the flux is
constant \citep{bottacini15}. To understand whether the entire integration time of this {\em NuSTAR} observation can be used for the spectral
analyses, we first explore the variability during the observation. Therefore, the light curve for both detectors is extracted binning the count rate
so that every bin lasts for 800 seconds. Such a bin size allows for detecting possible variability trends within the observation frame as in \cite{risaliti13}.
Figure \ref{fig:lc} displays the light curve of both modules, FPMA (red stars) and  FPMB (blue squares). For comparison, the dashed gray horizontal
line is the average count rate. The light curve shows no significant variability in accordance with \cite{oh18}. Being a Seyfert type-1 AGN, Mrk 876
allows for a rather unobscured view onto the innermost accretion region. In fact, high signal-to-noise {\em XMM}-Newton observations ($<$ 10 keV)
do not display any absorption in excess to the Galactic value \citep{porquet04, piconcelli05}, which excludes also any variability in column density.
Furthermore, observations performed by \cite{shull11} with the {\em Cosmic Origins Spectrograph} aboard the {\em Hubble Space Telescope},
confirm a very low column density towards the source. Thus, the spectral analyses capitalize on the entire observation time of {\em NuSTAR}.
\subsection{Spectral Analysis}
To confidently use $\chi$$^{2}$ statistics \citep{cash79, gehrels86}, {\em NuSTAR} data are binned to a minimum of 40 source counts bin$^{-1}$.
The data are fitted in {\texttt{XSPEC}} v 12.11.1 \citep{arnaud96} and errors are properly computed with {\texttt{XSPEC}}'s {\texttt{error}} command
at 1$\sigma$ level.
To all the spectral models used to fit the data, a constant component is added that allows for accounting for the cross-calibration of {\em NuSTAR}'s
detectors FPMA and FPMB by keeping either fixed to 1, while the other is free to vary. To account for the Galactic absorption towards Mrk~876,
assuming solar abundance, the rather precise measurement by \cite{elvis89} is used utilizing the NRAO 140 ft telescope of Green Bank finding a value
of N$_{H}^{gal}$ = 2.66 $\times$ 10$^{20}$ atoms cm$^{-2}$ with 5\% error, which is in good agreement with the averaged N$_{H}$ value from the
LAB Survey of Galactic H I \citep{kalberla05}.
\begin{figure}
\includegraphics[width=1.0\columnwidth]{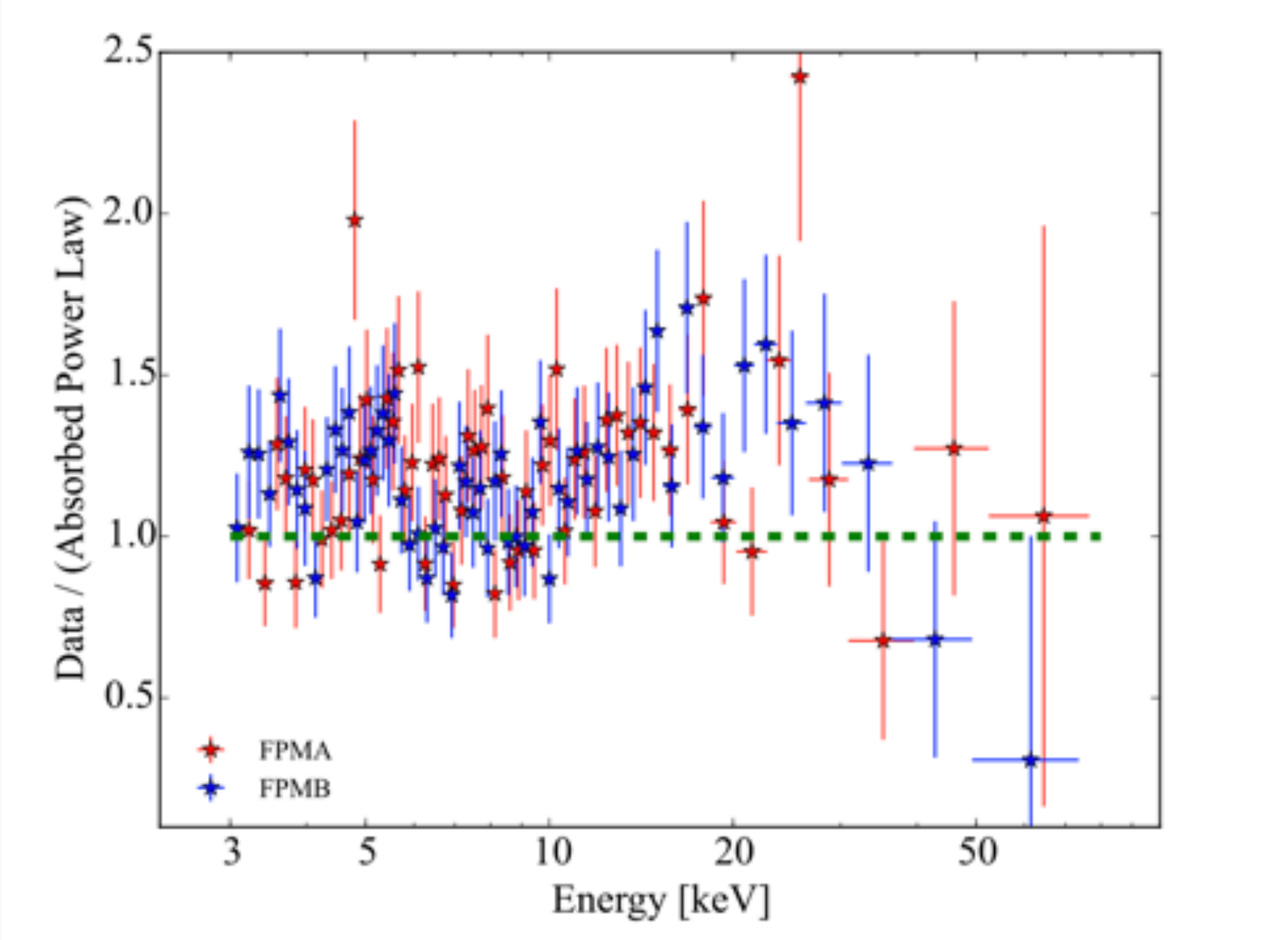}
\caption{Ratio between data and the folded model, which is an absorbed (fixed to the Galactic value) power-law
model (green horizontal line).}
\label{fig:ratio1}
\end{figure}
\begin{table}
	\centering
	\caption{Spectral parameters of the best fit model for {\em NuSTAR} spectrum of Mrk~876. 
Obs. ID: 60160633002, Obs. Date: 2020-10-22 UT 05:46:09, Obs Exposure: 29969}
	\label{tab:spectrum}
	\begin{tabular}{lcc}
		\hline
		Model Parameter & Value & Unit\\
		\hline
		\multicolumn{3}{c}{  \texttt{const(wabs(bknpower$+$gauss))} \vspace{2mm}} \\
\vspace{2mm}

C					 & 1.05$^{+0.01}_{-0.02}$ 			& ... \\

\vspace{2mm}

N$^{gal}_{H}$			& 2.6$\times$10$^{20}$ 				& $atoms~cm^{-2}$ \\

\vspace{2mm}

$\Gamma_{1}$			& 1.45$^{+0.16}_{-0.13}$ 				& ... \\

\vspace{2mm}

$\Gamma_{2}$			& 3.04$^{+1.76}_{-1.22}$ 				& ...	\\

\vspace{2mm}

Break Energy			& 25.95$^{+ 5.20}_{-12.82}$			& $keV$ \\

\vspace{2mm}

Norm (bknpower)		& 1.50$^{+1.25}_{-1.15}$	 			& $10^{-4}~ph~keV^{-1}~cm^{-2}~s^{-1}$  \\

\vspace{2mm}

Line energy			& 3.72$^{+1.42}_{-2.03}$		 		& $keV$ \\

\vspace{2mm}

Line width				& 1.42$^{+0.33}_{-0.31}$	 			& $keV$ \\

\vspace{2mm}

Norm (gauss)			& 8.18$^{+10.30}_{-6.80}$	 		& $10^{-5}~ph~s^{-1}~cm^{-2}$\\

\vspace{2mm}

Flux$_{3 - 79~keV}$		& 1.039$^{+0.176}_{-0.102}$	 		 & $10^{-12}~erg~cm^{-2}~s^{-1}$\\

\vspace{2mm}

$\chi^{2}$				& 115.54							& ... \\

\vspace{2mm}

d.o.f. 				& 114							& ...	\\
		\hline
	\end{tabular}
\end{table}

\begin{figure}
\includegraphics[width=1.0\columnwidth]{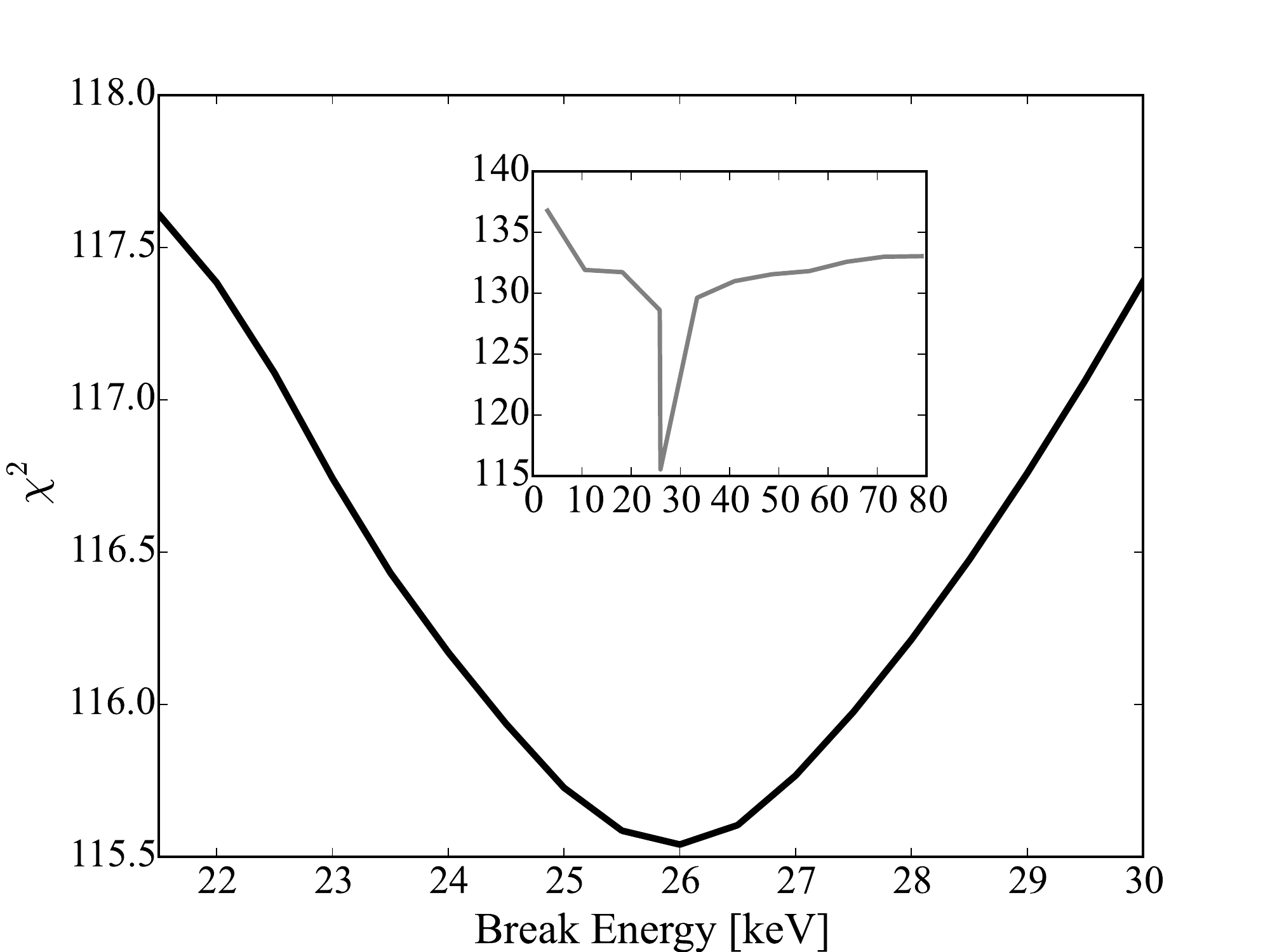}
\caption{Contour plot of the break energy. The inset shows the minimum to be global.}
\label{fig:steppar1}
\end{figure}

\subsubsection{The Broadband Spectral Analysis}
\begin{figure}
\includegraphics[width=1.0\columnwidth]{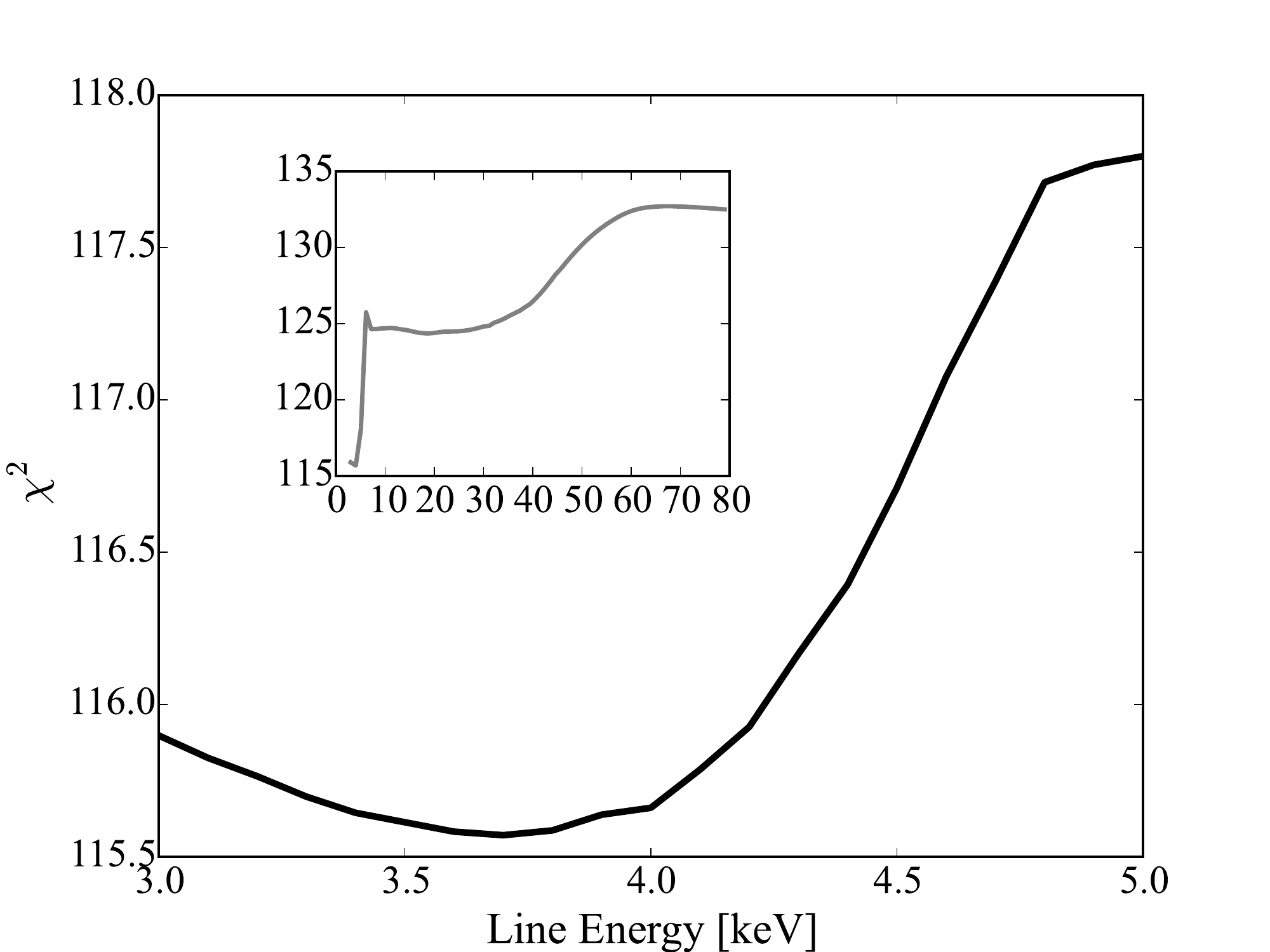}
\caption{Contour plot of the line energy. The inset shows the minimum to be global.}
\label{fig:steppar2}
\end{figure}

In a first attempt to fit the data, we adopt the simplest spectral shape, which is a power-law model having absorption fixed to the Galactic value
({\texttt{wabs*powerlaw}}). As a result the fit is unsatisfactory resulting in a reduced chi square of $\chi{^2}_{red}$=1.15. There is no evidence
for further absorption that could improve the fit, which is in agreement with previous studies performed using {\em XMM-Newton} and
{\em Swift}/XRT \citep{porquet04, piconcelli05, bottacini15}. The data-to-model ratio of this fit is shown in Figure~\ref{fig:ratio1}, which displays
excesses at energies between $\sim$3.5 -- 6.0 keV and between $\sim$10 -- 30 keV.
To improve the goodness of the fit the broadband is modeled with an absorbed broken
power-law model to mimic the excess at high energies. To model the excess between
$\sim$3.5 - 6.0 keV an additional broad Gaussian component is added to the model.
The complete model is given by an absorbed (fixed to the Galactic value) broken power law
plus a Gaussian component ({\texttt{wabs(bknpower+gauss)}}).
Except for the absorption that is fixed to the Galactic value, all parameters are free to vary.
As a result this leads to a satisfactory fit being the $\chi^{2}$=115.54 for 114 degrees of freedom.
The reduced chi square results in $\chi{^2}_{red}$=1.01. The complete fit results are reported
in Table~\ref{tab:spectrum}.
The break energy of the broken power law is well constrained at E$_{break}$=25.95 keV,
which is shown in Figure~\ref{fig:steppar1} where the break energy on the x-axis
constrains the best fit value for the smallest $\chi^{2}$ on the y-axis. 
The same type of analysis is shown in Figure~\ref{fig:steppar2} for the line energy of the Gaussian component,
which is also very well constrained. The insets in both, Figure~\ref{fig:steppar1} and Figure~\ref{fig:steppar2}, show that the obtained minimum
$\chi^{2}$ is global by scanning the entire parameter space.
\begin{figure}
\includegraphics[width=0.65\columnwidth,angle=-90]{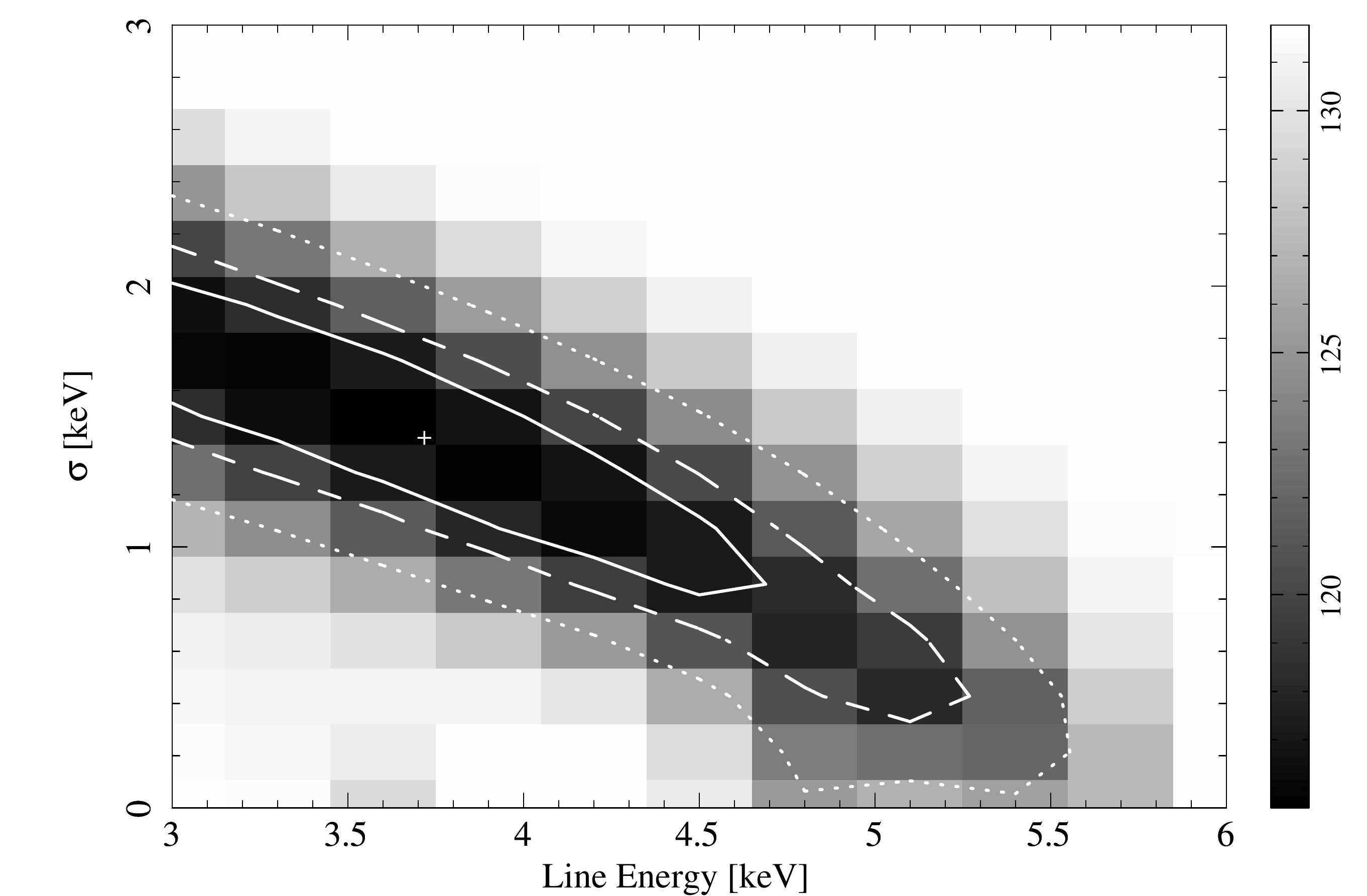}
\caption{Contours of the line energy (x-axis) and the line width $\sigma$ (y-axis): solid, dashed, and dotted lines delimit the
1$\sigma$, 2$\sigma$, and 3$\sigma$ levels, respectively. The vertical colour bar encodes the values of the $\chi^{2}$ seen
as background colour of the figure. The cross (+) at coordinates (3.72, 1.42) represents the best fit value.}
\label{fig:contours-bestfit}
\end{figure}

For the purpose of visualizing the fit parameters of the Gaussian component, in Figure~\ref{fig:contours-bestfit} we show
the contour levels of the width of the Gaussian (y-axis) as function of the centroid (x-axis). It is shown that both parameters are independently
constrained.\\
Even though the phenomenological {\texttt{wabs(bknpower+gauss)}} model precisely fits the data, also the more physically motivated 
{\texttt{wabs(cutoffpl+gauss)}} model is being explored. However, such a model worsens the $\chi{^2}_{red}$ resulting in 
$\chi{^2}_{red}$=1.20. By properly computing the fit errors and scanning the parameter spaces the fit improves but the normalization of the
Gaussian component remains unconstrained. Very importantly, the cutoff energy at 16.5 keV is found at energies too low for typical values
\cite[e.g.][]{balokovic20}.

\subsubsection{Reflection Features off the Accretion Disk and Relativistic Reflection Models}
As the Gaussian component might hint at the excess due to the joint effect of Doppler shift, relativistic beaming,
and gravitational redshift, its statistical significance needs to be established. There is a general agreement among scientists
that line searches driven by observational data must be validated through randomized trials  \citep{protassov02}. Therefore,
we perform Monte Carlo simulations to establish the probability whether the line be a statistical fluke or a real spectral feature.
This approach has been extensively used in many researches \citep[e.g.][]{turner10, tombesi10}.\\

\begin{figure}
\includegraphics[width=1.0\columnwidth]{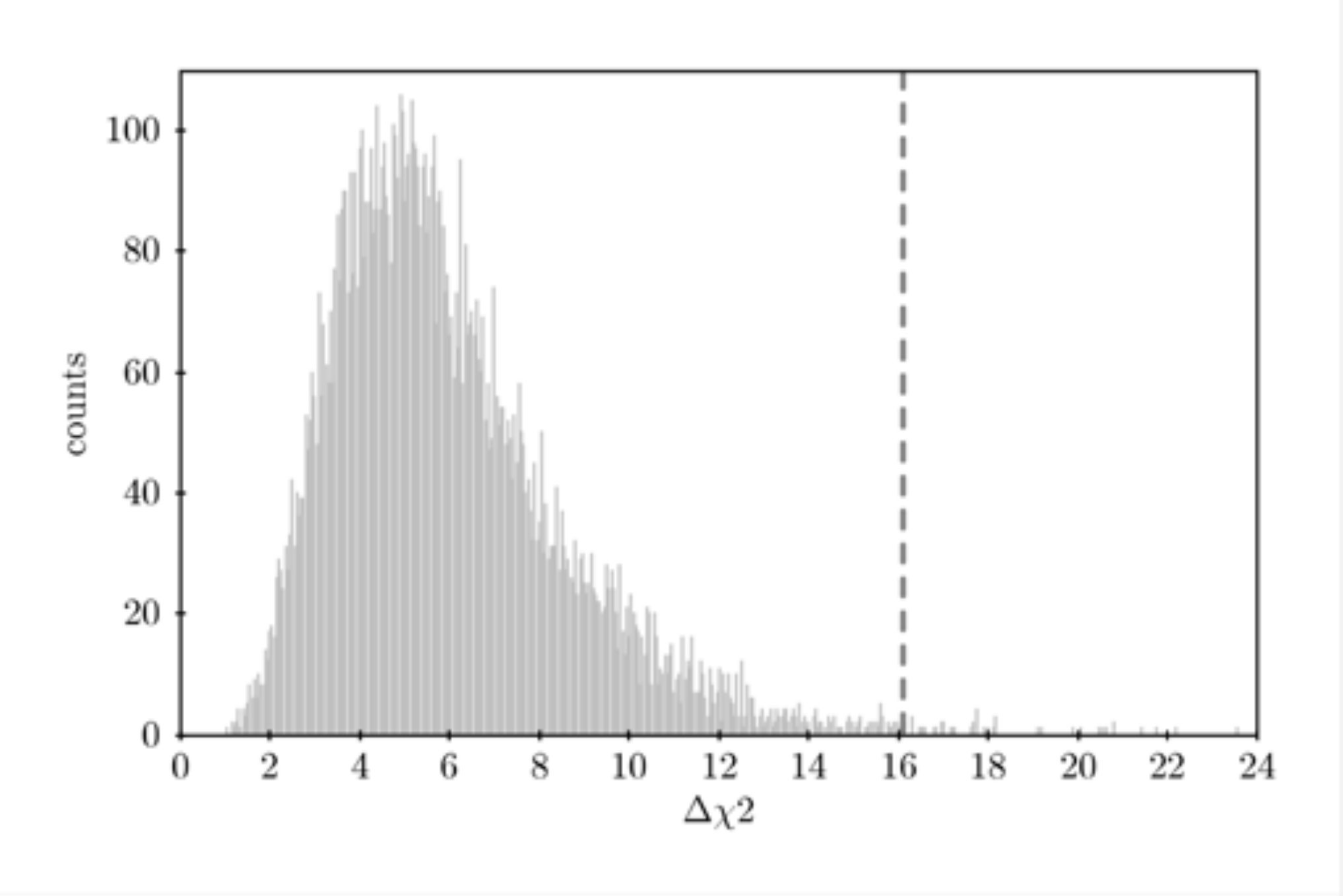}
\caption{Distribution of $\Delta \chi^{2}$ values obtained with the Monte Carlo simulation. Black dashed vertical
line represents the $\Delta \chi^{2}_{thr}$ derived from the actual observation. Three occurrences exceed this value
in 10$^{4}$ trials in this simulation thereby rejecting the null-hypothesis at a significance level of $\sim$3$\sigma$ (99.71\%),
which excludes the broad spectral excess between $\sim$3.5 -- 6.0 keV to be a statistical fluke.}
\label{fig:monte-carlo}
\end{figure}

\begin{table*}
\begin{center}
\caption{Spectral fit results of twelve {\em Chandra} observations between 2016 and 2017 using the best fit model {\texttt{wabs*bknpower}}. Fluxes are reported in the range 1-7 keV.}
\label{tab:chandra-spec}
\begin{tabular}{ccccllcccc}
\hline
\hline 
Chandra & Start  & Exposure & $\Gamma$$_{1}$ & E$_{brk}$ &  $\Gamma$$_{2}$     & Norm                                                                & $\chi$$^{2}$ & d.o.f. & Flux$_{~1 - 7~keV}$ \\
obs id     & [date] & [s]            &                              &  [keV]        &                                   & [10$^{-3}$ ph keV$^{-1}$ cm$^{-2}$ s$^{-1}$] &                      &          & [10$^{-12}$ erg~cm$^{-2}$ s$^{-1}$]  \\
 \hline
\\
\vspace{1mm}
18144 & 2016-03-18T04:18:40 & 23920 & 2.10$^{2.18}_{2.02}$ &  4.36$^{4.86}_{3.99}$ &  0.20$^{0.73}_{-0.51}$ & 1.67$^{1.77}_{1.56}$ & 129.73 & 131 & 5.26$^{5.36}_{5.15}$ \\
\vspace{1mm}
18145 & 2016-03-19T22:13:59 & 23100 & 2.57$^{2.87}_{2.35}$ &  1.60$^{1.81}_{1.37}$ &  1.69$^{1.79}_{1.59}$ & 1.78$^{1.91}_{1.59}$ & 110.17 & 105 & 5.18$^{5.43}_{5.09}$ \\
\vspace{1mm}
18146 & 2016-05-06T06:01:40 & 22950 & 2.39$^{2.59}_{2.21}$ &  1.75$^{1.98}_{1.54}$ &  1.61$^{1.71}_{1.51}$ & 1.79$^{1.90}_{1.68}$ & 120.38 & 112 & 5.58$^{5.83}_{5.52}$ \\
\vspace{1mm}
18147 & 2016-09-24T16:16:51& 11950 & 2.67$^{3.04}_{2.33}$ &  1.65$^{2.18}_{1.39}$ &  1.57$^{1.70}_{1.43}$ & 1.67$^{1.43}_{1.70}$ & 62.94 & 66 & 6.35$^{6.75}_{6.24}$ \\
\vspace{1mm}
18148 & 2017-04-02T03:18:42 & 17170 & 2.64$^{2.90}_{2.35}$ &  1.59$^{1.88}_{1.38}$ &  1.65$^{1.77}_{1.53}$ & 1.60$^{1.73}_{1.46}$ & 70.93 & 68 & 4.63$^{4.91}_{4.54}$ \\
\vspace{1mm}
18799 & 2016-03-21T00:07:04 & 24950 & 2.72$^{2.95}_{2.48}$ &  1.54$^{1.79}_{1.40}$ &  1.63$^{1.71}_{1.54}$ & 1.90$^{2.01}_{1.79}$ & 124.61 & 118 & 5.58$^{5.78}_{5.50}$ \\
\vspace{1mm}
18814 & 2016-03-18T23:14:21 & 23930 & 2.35$^{2.66}_{2.24}$ &  2.63$^{1.37}_{3.04}$ &  1.55$^{1.84}_{1.36}$ & 2.03$^{2.15}_{1.91}$ & 112.94 & 110 & 5.44$^{5.58}_{5.34}$ \\
\vspace{1mm}
18844 & 2016-05-07T09:28:25 & 22950 & 2.54$^{2.78}_{2.35}$ &  1.76$^{1.98}_{1.53}$ &  1.59$^{1.70}_{1.49}$ & 1.81$^{1.93}_{1.70}$ & 77.48 & 102 & 5.30$^{5.49}_{5.21}$ \\
\vspace{1mm}
19885 & 2016-09-21T01:16:27 & 25660 & 2.59$^{2.86}_{2.38}$ &  1.62$^{1.87}_{1.39}$ &  1.67$^{1.75}_{1.58}$ & 2.31$^{2.43}_{2.18}$ & 135.84 & 136 & 6.68$^{6.91}_{6.61}$ \\
\vspace{1mm}
19889 & 2016-09-25T20:14:49 & 12530 & 2.06$^{2.15}_{1.97}$ &  4.44$^{5.07}_{4.02}$ &  0.63$^{1.17}_{-0.19}$ & 2.13$^{2.29}_{1.97}$ & 89.36 & 96 & 6.71$^{6.87}_{6.51}$ \\
\vspace{1mm}
20022 & 2017-03-28T17:08:24 & 22840 & 2.73$^{3.01}_{2.39}$ &  1.48$^{1.93}_{1.35}$ &  1.71$^{1.81}_{1.60}$ & 1.74$^{1.87}_{1.62}$ & 103.33 & 94 & 4.91$^{5.20}_{4.87}$ \\
\vspace{1mm}
20047 & 2017-03-29T06:15:47 & 9940 & 2.98$^{3.95}_{2.10}$ &  1.28$^{2.07}_{1.08}$ &  1.74$^{1.86}_{1.56}$ & 1.63$^{1.83}_{1.42}$ & 40.42 & 41 & 4.86$^{5.90}_{4.80}$ \\  
\\                  
\hline
\hline
\label{Table1}
\end{tabular}
\end{center}
\end{table*}

To establish the significance of the rather broad excess, Monte Carlo simulations in {\em NuSTAR}'s entire energy range
between 3 -- 79 keV are performed. The null hypothesis for this simulation is: the spectrum measured by {\em NuSTAR} is
a broken power law with absorption fixed to the Galactic value (null model). This spectrum is simulated with {\texttt{XSPEC}}
using the {\texttt{fakeit}} command accounting for the instrument's response files of the actual observation and its exposure.
The resulting simulated spectral data are grouped to a minimum of 40 counts bin$^{-1}$ as is the measured spectrum of the
actual observation. Such a binning allows for an adequate statistics and the confident use of $\chi$$^{2}$ statistics \citep{cash79}.
This procedure is iterated 10$^{4}$
times. Each of the spectra are fitted with the null model. The results are used to simulate a
further spectrum in the very same way as performed before. This allows for accounting for the uncertainties of the null hypothesis
model as described by \cite{markowitz06} and by \cite{tombesi10}. Each obtained simulated spectrum is again fitted with the null
model obtaining the $\chi^{2}_{null}$. This will be the reference $\chi^{2}$. The very same simulated spectrum is then fitted with 
the null model and an additional Gaussian component throughout the range 3 -- 79 keV. This energy band is stepped through
by energy bins, whose size corresponds to the actual best spectral energy resolution of 0.4 keV of {\em NuSTAR} \citep{harrison13}.
Thus, the simulations will lead to a rather conservative result. During the single fit procedures all parameters are free to float.
Especially the normalization value of the Gaussian component varies freely between both, negative and positive values. For each
simulated spectrum the best chi square ($\chi^{2}_{best}$) is used to maximize $\Delta \chi^{2}$ (= $\chi^{2}_{best}$ - $\chi^{2}_{null}$).
Thus, the distribution of the 10$^{4}$ $\Delta \chi^{2}$ determines the fraction of lines caused by chance fluctuations whenever the value
of $\Delta \chi^{2}$ exceeds the threshold value of $\Delta \chi^{2}_{thr}$= 16.09 (131.63 - 115.54) obtained from the actual observation
(131.63 and 115.54 being the two $\chi^{2}$ without and with the Gaussian respectively).
As a result 29 out of the 10$^{4}$
$\Delta \chi^{2}$ exceed the value of $\Delta \chi^{2}_{thr}$. Therefore, the null hypothesis of the
measured spectrum being a broken power-law model with Galactic absorption is rejected with a probability of
99.71\%, which corresponds to $\sim$3$\sigma$. This result excludes the broad Gaussian component to be a statistical fluke.
The distribution of the 10$^{4}$
$\Delta \chi^{2}$ values is displayed in Figure~\ref{fig:monte-carlo}, where the black dashed vertical line represents
$\Delta \chi^{2}_{thr}$.
Since the modeled excess between $\sim$3.5 -- 6.0 keV and the 
modeled break energy at 25.95 keV improve the fit, we study these rather strong emission features that are typical for relativistic
reflection off the accretion disk surrounding the rotating supermassive black hole.
These features might be due to the blurring at the accretion disk caused by the strong Doppler and gravitational shifts and by the
gravitational redshift. These effects are convolved with the rest-frame X-ray reflection in the widely used self-consistent model
{\texttt{RELXILL}} \citep[version v1.4.3;][]{garcia14, dauser14}.
This model selfconsiestently accounts
for the relativistic reflection physics at work in the vicinity of a black hole. Also it assumes the accretion disk to be
irradiated by a power-law coronal emitter. The accretion disk reprocesses the radiation through several steps including the
gravitational redshift and the relativistic broadening. By fitting this model to the data, it allows for estimating the accretion disk
ionization ($\xi$), the fraction of reflected radiation ($R$), the iron abundance ($A_{Fe}$), the inclination angel of the accretion
disk ($\theta$), and the dimensionless spin ($a$), which is allowed to assume values -1~$<$~$a$~$<$~1 for retrograde and
prograde motion. Given the rather short observing time for such an analysis, we approximate {\texttt{RELXILL}}'s primary
radiation to a simple power-law (rather than a broken power law) by equalizing the emissivity index above and below the
break making the break itself an unused parameter in this fit.
The inferred emissivity index is $q=~4.56^{+0.70}_{-0.98}$. The spin results in an upper limit of $a~\leq~0.85$,
while the inclination angle has a rather large error range of $i=32.84$\degree{}$^{+12.22}_{-8.99}$, which however is
supported by the inclination angle inferred by \cite{bian02}. For the accretion disk ionization state an upper limit of
log($\xi$)~$\leq$~3.17 is inferred, while the iron abundance A$_{Fe}$=~1.85$^{+2.36}_{-1.24}$ is nearly twice the
solar value. The reflection fraction has a rather large uncertainty $1.91^{+17.54}_{-1.13}$. The inner radius of the
accretion disk in unconstrained, while the outer radius is fixed to 400 gravitational radii.\\ 
In the following we also explore the reflection of the primary radiation off material, which is at greater distance
from the SMBH. This greater distance from the gravitational potential allows for modeling the spectra
without the need of convolving the radiation with relativistic effects. To account for this reflection, we fit the 
spectra with the {\texttt{xillver}} reflection model \citep{garcia10, garcia13} in addition to the {\texttt{cutoffpl}} model
to account for the primary radiation. The resulting model is {\texttt{wabs(cutoffpl+xillver)}}. While fitting, the
parameters are free to vary and the cut-off energy of the {\texttt{cutoffpl}} is fixed to the same parameter
of the {\texttt{xillver}} component. The fitted value for this parameter is E$_{cutoff}$=~11.59$^{+12.32}_{-9.06}$,
which coincides with the Compton hump. Such cut-off energy is much smaller compared to values routinely found
in AGN at energies $\sim$ 200 keV. The inclination angle $i$ and the normalization of the {\texttt{xillver}} component remain unconstrained.\\
Additionally we explore the {\texttt{pexmon}} model \citep{nandra07}, which includes also self-consistent lines of Fe K$_{\alpha},$
Fe K$_{\beta}$ and the Compton shoulder \citep{george91}. A further {\texttt{cutoffpl}} is added. Also for this reflection fit all
the parameters are free to float. The photon index of the {\texttt{pexmon}} component is tied to the same parameter of the
{\texttt{cutoffpl}} component. Also the two cut-off energies of the two model components are tied. For this model the inclination
angle $i$ and the iron abundance A$_{Fe}$ are unconstrained, while for the cut-off energy a lower limit of E$_{cutoff}$>0.01 keV
is derived.
 
\subsection{X-ray Observations Below 10 keV}
\subsubsection{Swift/XRT and XMM-Newton Observations} 
Observations of Mrk~876 by {\em Swift}/XRT and {\em XMM-Newton} have been presented in \cite{bottacini15}. 
{\em Swift}/XRT observations are able to constrain a simple absorbed power-law model only, except for one observation,
which displays a transient Fe line at 99\% probability. This transient and gravitationally redshifted line originates in a short-lived
hotspot on the accretion disk that is due to a magnetic reconnection event in the corona. On the other hand the two {\em XMM-Newton}
observations are able to constrain a slightly more sophisticated absorbed broken power-law model. The break energy at
$\sim$ 1.8 keV and the rise of the spectra toward lower energies might hint at some soft excess, which is in agreement
with the two independent analyses by \cite{porquet04} and \cite{piconcelli05}. None of the mentioned analyses including
the analysis in \cite{bottacini15} detect an Fe line.
\subsubsection{Chandra Observations} 
\begin{figure}
\includegraphics[width=0.65\columnwidth,angle=-90]{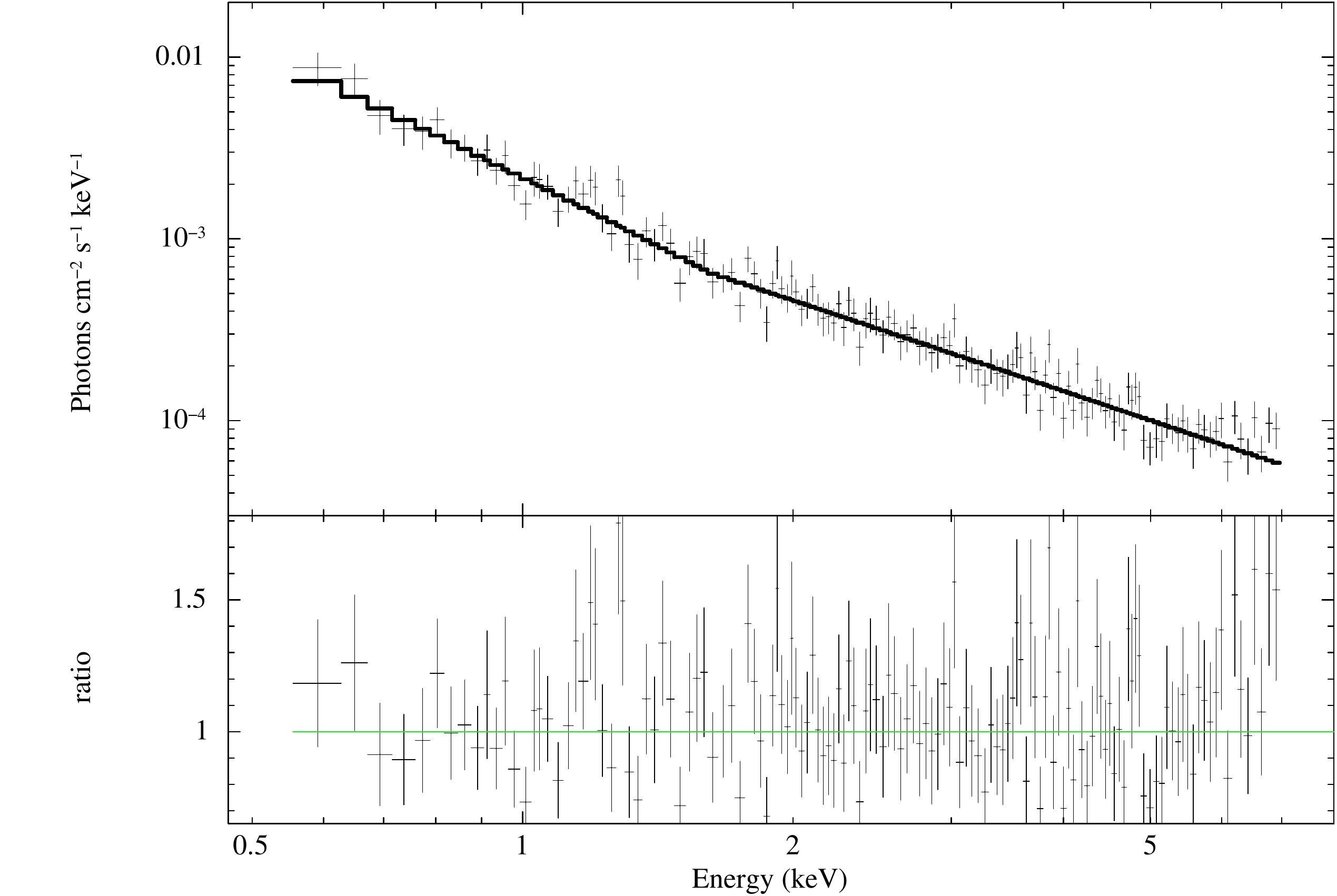}
\caption{Top panel: Absorbed (fixed to the Galactic value) broken power-law model for {\em Chandra} observation 19885. Bottom
panel: ratio between data and model.
\label{fig:chandraspec}}
\end{figure}

\begin{figure}
\includegraphics[width=1.0\columnwidth]{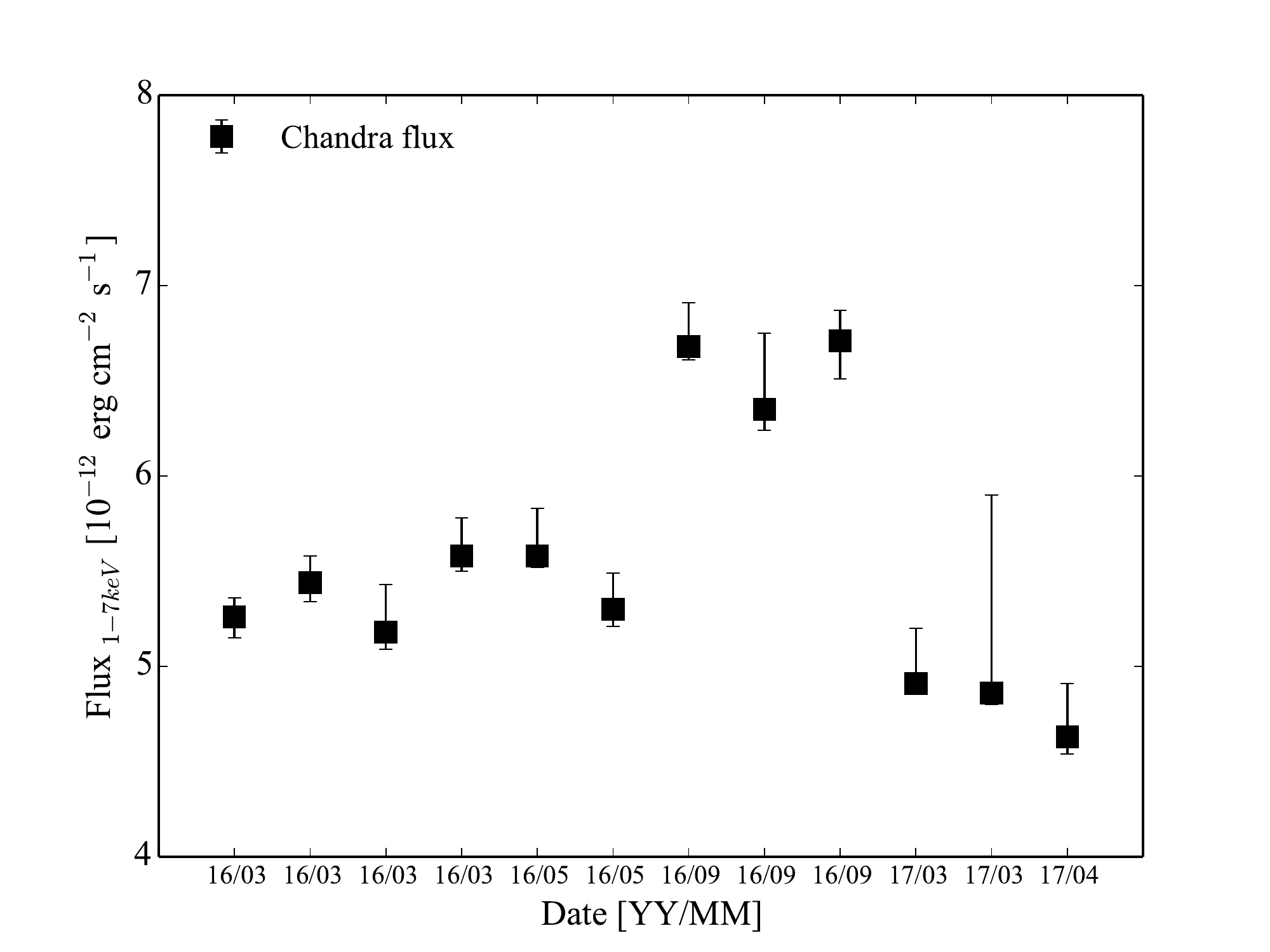}
\caption{{\em Chandra} light curve of Mrk~876 as observed from 2016 to 2017. 
\label{fig:chandralc}}
\end{figure}

The {\em Chandra} X-ray Observatory has been monitoring Mrk~876 in the astrophysical context of AGN feedback through outflows
with the Advanced CCD Imaging Spectrometer spectroscopic array (ACIS-S) from 2016 to 2017 with 12 observations. These observations
have an average exposure time of $\sim$ 20 ksec.
Dataset and relevant information can be found in Table~\ref{tab:chandra-spec}. These {\em Chandra} observations were carried out in FAINT mode. Data
have been analyzed using the standard {{\texttt{chandra\_repro}} script in CIAO \citep{fruscione06} data analysis system v. 4.13
and {\em Chandra} calibration database CALDB v. 4.9.4. Spectra have been extracted with the {\texttt{specextract}} tool. Additionally
they have been binned to contain a minimum of 20 counts bin$^{-1}$. Errors are reported at 1$\sigma$ level. All the spectra are best fitted
with an absorbed (fixed to the Galactic value) broken power-law model. The break at $\sim$1.6 keV hints at a soft excess, which is consistent with
the findings by the observations with {\em XMM-Newton}. Figure~\ref{fig:chandraspec}  shows the spectrum of observation id 19885 having
a break energy of E$_{brk}$=1.6 keV. Observations 18144 and 19889 display a larger break energy of $\sim$4.4 keV.
None of the {\em Chandra} observations need absorption in excess to the Galactic value, neither at the source nor along the line of sight.
This is in agreement with the previous spectral fit results by {\em XMM}-Newton and {\em Swift}/XRT. {\em Chandra} observations
detect flux variability that can be pictured in Figure~\ref{fig:chandralc}. The amplitude of the variability is of 40\%. Such variability
had not been detected in previous X-ray measurements. 
While the High-Energy Transmission Grating (HETG) spectrometer with its preferred ACIS-S array (used for these observations)
has the ability to spectrally resolve high-velocity outflows and narrow atomic lines, it also is limited to bright sources. Mrk~876 exhibits
a moderate flux (average flux in the 1--7 keV band 5.5$\times$10$^{12}$ erg cm$^{-2}$ s$^{-1}$ see Table~\ref{tab:chandra-spec}),
which might prevent from detecting sharp spectral lines. Neither the very similar observations by {\em XMM-Newton} are able to detect
sharp spectral features even though its effective area is much larger especially at high energies.
\begin{figure}
\includegraphics[width=0.7\columnwidth,angle=-90]{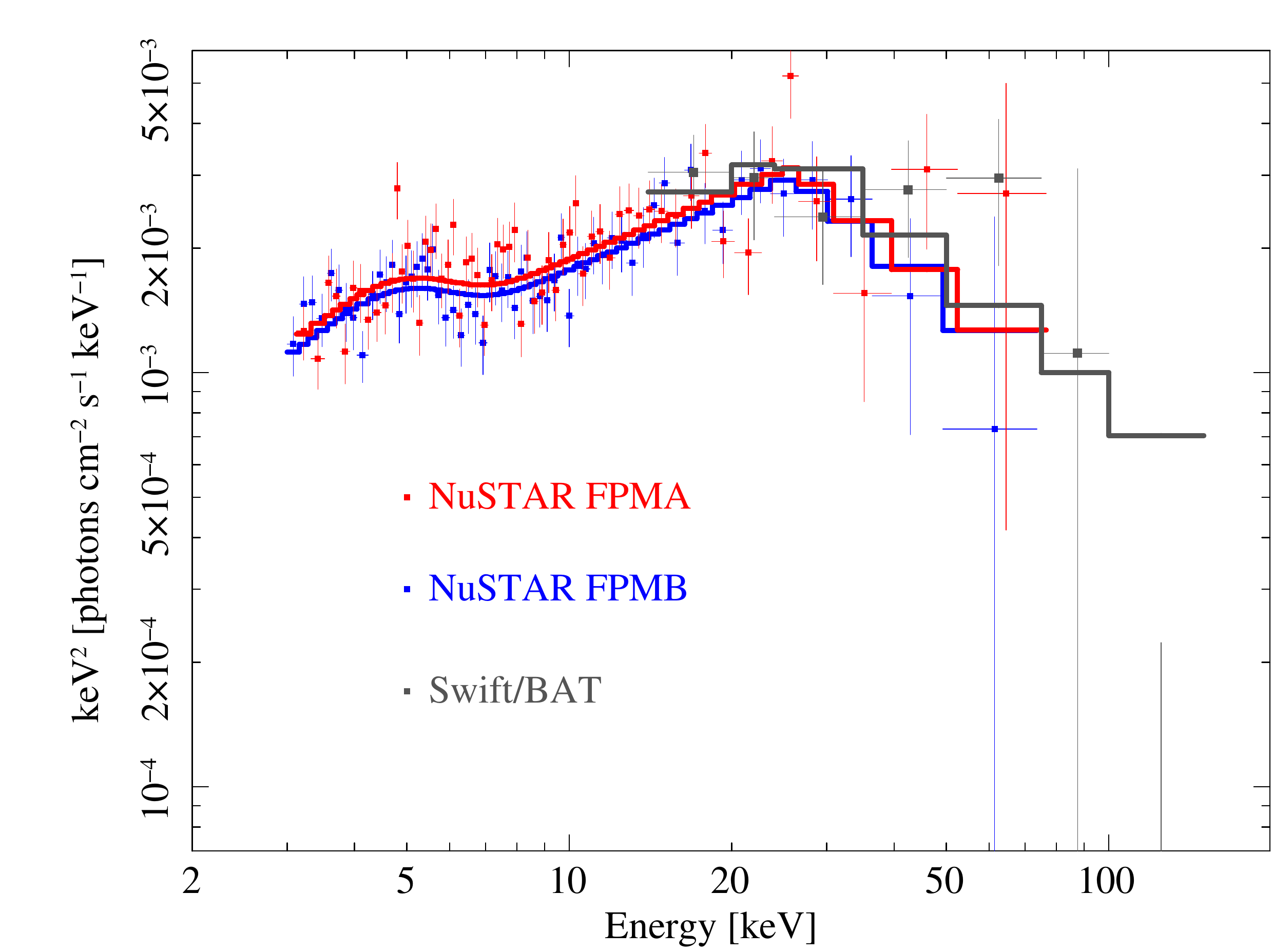}
\caption{Combined spectra of {\em NuSTAR}'s FMPA, FMPB, and {\em Swift}/BAT.
\label{fig:nustar-bat}}
\end{figure}

\subsection{Discussion}
As the analyzed data in this research show that Mrk~876 is variable, the combined use of soft X-ray data and {\em NuSTAR}
data would require simultaneous observations that were actually not performed. However, we compare the flux in the overlapping energy
range 3 - 7 keV of the most recent {\em Chandra} and {\em NuSTAR} observations. The closest observations in time between the two missions
({\em Chandra} obs id 18148, which is still $\sim$ 3 years apart) shows the lowest {\em Chandra} flux level of 2.4$\times$10$^{-12}$ erg cm$^{-2}$ s$^{-1}$.
Yet, this level is more than 20\% higher than the flux measured by {\em NuSTAR} (1.9$\times$10$^{-12}$ erg cm$^{-2}$ s$^{-1}$), that prevents
from fitting the combined datasets.\\
At the high-end range of the spectrum, the cut-off energy of the continuum emission is routinely found well above 30 keV, at 
which energy the transfer to the photons is not efficient anymore and the spectrum falls off. This is observationally confirmed by
detailed spectroscopic analyses of the {\em Swift}/BAT sample that displays a median measured cut-off energy of 76 $\pm$ 6 keV
\citep{ricci17}. However the same study shows that when accounting also for the lower and upper limits derived from the same sample,
the median cut-off energy is even higher at 200 $\pm$ 29 keV.
This result has been confirmed very recently with a sample study by \cite{balokovic20} with the more sensitive {\em NuSTAR}
measurements.
Therefore, it is possible to exclude the spectral turnover at 25.95 keV
to be associated to the cut off by the continuum. This becomes apparent from Figure~\ref{fig:nustar-bat} that shows the jointly fitted
{\em NuSTAR} and {\em Swift}/BAT \citep{oh18} spectra with the broken power law plus Gaussian model. BAT data (dark gray markers)
extend to somewhat higher energies without detecting the cut-off spectral feature and the break at 25.95 keV represents the data well.
Through Monte Carlo simulations, the excess at energies $\sim$3.5 -- 6.0 keV has been proven statistically significant to
99.71\% (3$\sigma$).
This spectral feature and the spectral break at
25.95 keV tie in well with the relativistic disk reflection scenario \citep[][for a very recent review]{reynolds21}. As an alternative
scenario we explore also the warm absorber hypothesis even though Mrk~876's spectra do not exhibit strong absorption
edges. Therefore we fit in addition to the absorbed broken power-law model a warm absorber ({\texttt{zxipcf}) \citep{reeves08}.
Such a model would mimic partially ionized absorbing matter that partially intercepts the primary continuum from the source along
the line of sight to the observer. This yields a rather good (although not best) fit result ($\chi{^2}_{red}$=1.05) even thought the
ionization parameter log($\xi$)~$=$~3.0 erg s$^{-1}$ and the associated column density N$_{H}$~$=$~10$^{24}$~cm$^{-2}$ are
physically unsatisfactory values for such intervening matter \citep[][and references therein]{tombesi13}. Both are inconsistent with
typical values in the range log($\xi$)~$\sim$~0 -- 2 erg s$^{-1}$ and N$_{H}$~$\sim$~10$^{20}$ - 10$^{22}$ atoms cm$^{-2}$
\citep{McKernan07, tombesi13}. Especially this latter value is well established and in good agreement with a recent dynamical
modeling of warm absorber motion in AGN \citep{kallman19}. Such models predict values of N$_{H}$ to be at most 10$^{22}$~cm$^{-2}$
for small viewing angles as for Mrk~876. Additionally, and very importantly, the warm absorber scenario is unable to account for the
spectral break at 25.95 keV. A further issue for the warm absorber scenario comes from the result of the fitted values themselves. In fact,
if intervening matter of N$_{H}$~$\sim$~10$^{24}$~cm$^{-2}$ would cross the line of sight, then the inevitable efficient absorption would
lead to detectable variability on short time scales as warm absorbers are found at velocities $v$~$>$~100 km s$^{-1}$ \citep{laha14} up
to mild relativistic velocities \cite[e.g. $\sim$0.3c;][]{braito07}. Such variability has never been detected in many observations for
Mrk~876 including the present {\em NuSTAR} and {\em Chandra} observations of this research. Also the neutral Compton reflection off clumpy and
optically thick distant clouds would imply absorption variability \citep{turner00, miller08} never observed in Mrk~876.\\
We also study the distant reflection scenario, whose distinct feature would be a narrow (rather than a large) Gaussian component.
To explore the distant reflection scenario the combined use of the {\texttt{cutoffpl}} and {\texttt{xillver}} components are unable
to constrain the inclination angle $i$. Also the inferred value of the cut-off energy (E$_{cutoff}$=11.59 keV) is much lower than the typical
values inferred through observations \citep[e.g.][]{ricci17, balokovic20}. However, this does not come as a surprise. In fact, the {\texttt{cutoffpl}}
in addition to a Gaussian is unable to properly fit the spectra ($\chi_{red}^{2}$=1.20, see section 2.3.1). The same limitations
apply to the  {\texttt{cutoffpl+pexmon}} model. Furthermore, also this model does not allow for constraining the inclination angle $i$.\\
As very recently discussed by the authors of \cite{kamraj22}, measurements of the cut-off energy should be taken with a skeptical
attitude because the values of such measurements are largely affected by the uncertainties in modeling the data and by the quality of
the data. Even more so, a such low cut-off energy as inferred in our distant reflection is not measured by current most sensitive
observations with {\em NuSTAR} (for a sample study see \cite{kang22} and for a case study see \cite{balokovic21}) nor is it expected
theoretically. From a theoretical view, the inverse-Compton scattering of thermal seed photons arising from the accretion disk by the
energetic particles of the corona is described by the Kompaneets equation  \citep{kompaneets57}. Its solution by \cite{lightman87}
and by \cite{zdziarski96} predicts a power-law spectrum $N(E)=KE$$^{-\gamma}$, for which the spectral index is given by:
\begin{equation}
\alpha=\sqrt{{9\over 4} +{1\over (kT/m_e c^2)\tau(1+\tau/3)}} - {3\over 2}
\label{Gamma}
\end{equation}
where $T$ is the coronal temperature and $\tau$ is the optical depth to the Thomson scattering. This equation holds true for X-ray continua at 
photon energies much less than the coronal temperature. For higher photon energies the spectrum falls abruptly off, because the electrons
of the corona are not energetic enough to up-scatter a large number of photons. Since hard X-ray surveys \citep[e.g.][]{vasudevan13} find 
temperatures well above 100 keV, much lower cut-off measurements as for our distant reflection model are not justified.
To further explore the cut-off energy for the
distant reflection, we model this emission scenario by applying a similar value of the reflection fraction as inferred with the {\texttt{RELXILL}}
model ($\Gamma$$\sim$2). This value affects mostly the high-energy component of the spectrum and thereby also the cut-off energy. Indeed, by
modeling the distant reflection ({\texttt{cutoff+xillver}}) with this approach the {\texttt{xillver}} component dominates over the {\texttt{cutoff}}
component. The normalization of this latter component cannot be constrained by the fit and the value of the cut-off energy shifts to lower
values of E$_{cutoff}$$\sim$9 keV. Therefore, even by enhancing the contribution of the reflection component, the cut-off energy cannot be
found at values that are justified by theory or observations.
On the other hand, the relativistic reflection scenario is able to naturally reproduce the broadband spectral features observed
with {\em NuSTAR} and shown in Figure~\ref{fig:seds}. These spectral features largely depend on the spin of the central
supermassive black hole. In Mrk~876 its spin was already put forward \citep{bottacini15} in the context of the transient hotspot
scenario \citep{nayakshin01, turner02} very similar to a precise time-resolved study by \cite{nardini16}.\\
Given the above evidences, it is reasonable to fit Mrk~876's {\em NuSTAR} spectra
with a more sophisticated model that incorporates the physics of the accretion disk in the vicinity of the black hole,
even though the relativistic reflection features are already mimicked and fitted by using the model given by the sum of the broken
power law and the large Gaussian component for the low energy part. These features can be reproduced by fitting the
{\texttt{RELXILL}} model that accounts for the entire
energy band from 3 -- 79 keV.  This model constrains the spin to an upper limit of $a~\leq~0.85$ and the inclination angle
of the disk is $i$=32.84\degree{}$^{+12.22}_{-8.99}$. The inclination angle is in agreement with
an independent measurements by \cite{bian02}. By using this more sophisticated spectral model we note that some
parameters remain unconstrained because of the well known degeneracies among parameters of such models
\citep{reynolds21} and because of the moderate signal-to-noise observation (for such analyses), which however is
enough to detect the actual relativistic reflection features.
The properly computed uncertainties cannot be significantly reduced by fixing parameters during the fitting process.
Therefore, also any discussion to better comprehend the physics at work in the environment of the emission region
would lead to speculative conclusions.
It is noteworthy that the inferred index of the emissivity profile, which was approximated to a
simple power law, is rather steep ($\sim4$). Such steep values are
in agreement with theoretical results related to the effects by gravitational light bending in the vicinity of spinning black
holes \citep{miniutti03}. \\
\begin{figure}
\includegraphics[width=1.0\columnwidth]{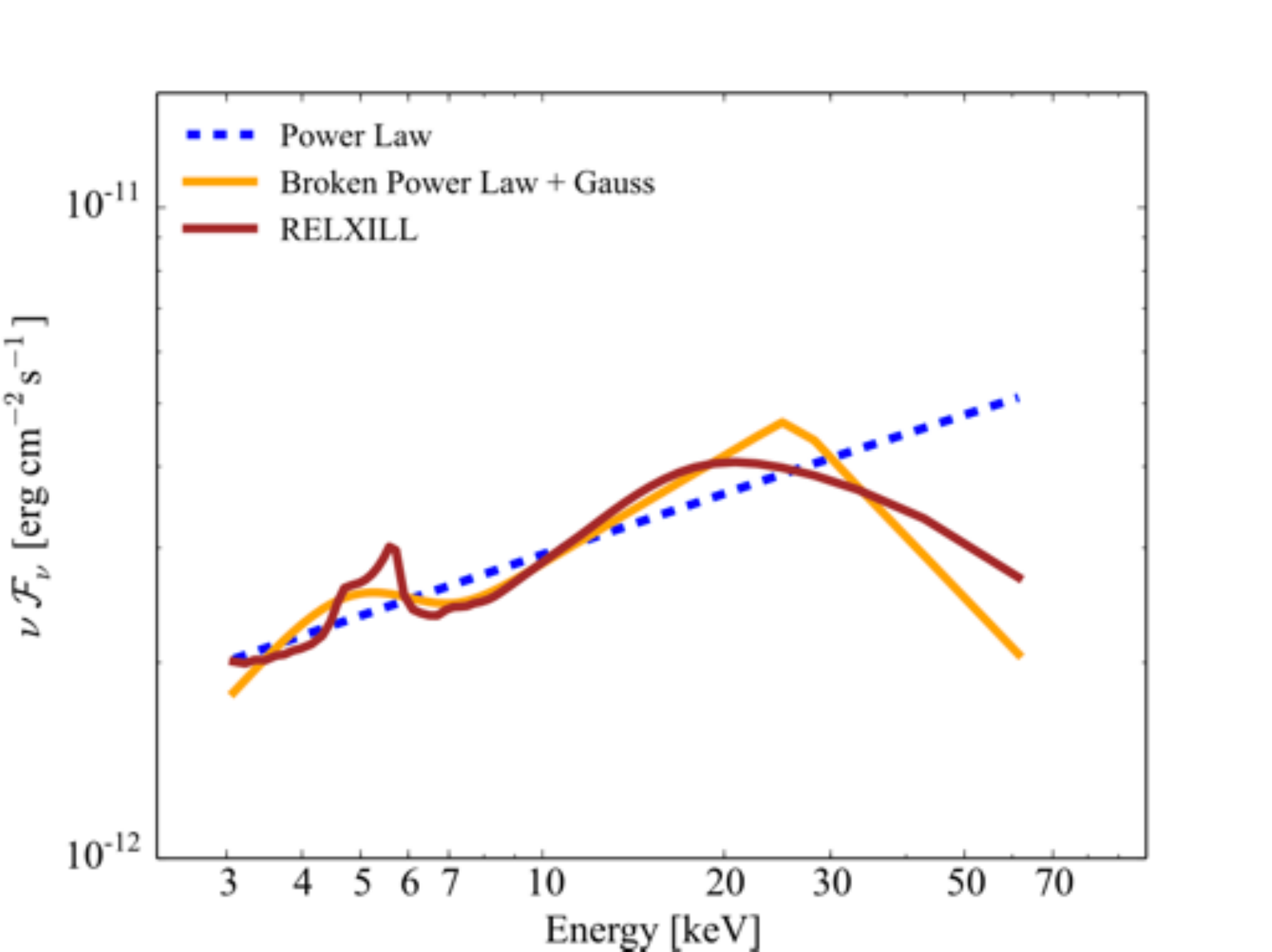}
\caption{SEDs in form of $E^{2}$$\times$$f(E)$ of the best fit model (broken power law + Gaussian, orange), {\texttt{RELXILL}} (brown), and
simple {\texttt{power law}} (dashed blue).
\label{fig:seds}}
\end{figure}

Figure~\ref{fig:ratios} shows a comparison of the residuals obtained through the fit results using the previously discussed models.
The top panel displays the residuals with respect to the simple power-law fit. The second panel from the top shows that the
sum of the broken power law and the Gaussian is able to resemble well the overall broadband spectrum. However, it
leaves some excess residuals between 5 -- 6 keV behind. In this interval the excess rises towards increasing 
energies only to fall off abruptly. In the view of the rotating black hole this feature could hint at the red wing of the accretion
disk. This feature is largely mitigated in the third panel
({\texttt{RELXILL}}) from the top. Indeed, Figure~\ref{fig:seds} displays the two models and also the simple power-law model
for comparison. It is apparent that even though the self-consistent reflection model (solid brown) is affected by some statistically
unconstrained parameters, it is able to reproduce the best fit model (solid orange). These two models
depart from the pure power-law model (dashed blue) to resemble the break at $\sim$26 keV and the excess between 
$\sim$3.5 -- 6.0 keV. This latter feature aligns with the cosmologically redshifted Fe~K-shell line system in the rest frame band
6.4 -- 6.97 keV. This line system tends to a noticeable red wing, which is the extent of the line to low energies,
because the inclination angle of the accretion disk (with respect to the line of sight) is modest \citep[$\sim30\degree{}$;][]{fabian00, reynolds21}.
This can be pictured in the spectral residuals in Figure~\ref{fig:ratios} first two panels from top. The extent to lower energies
decays smoothly. To further investigate this spectral feature, a broken power-law model plus a narrow (rather than a broad)
Gaussian line to reproduce the excesses of the residuals is being fitted. By properly tuning the initial line energy
($\sim$5.5 keV), the resulting fit leads to a good $\chi_{red}^{2}$=1.12. However, when computing the errors, the fit clearly 
does not allow for constraining the parameters of the line, which hints at the fact that the line might be broadened and skewed
through the joined action of the relativistic Doppler shift and the relativistic beaming.
We examine the parameter space of the line width and its energy (i.e. position of the centroid) calculating the
corresponding $\chi^{2}$. We show the contours of these two line parameters in Figure~\ref{fig:contours} where the dark
shaded area encodes low $\chi^{2}$ values for larger values of the line width and for smaller values than 5.5 keV of the
line energy. Therefore, the line is allowed to shift towards lower energies and the line width can become larger only to
obtain equally good fit results. This highlights that the complex Fe-line system decays smoothly towards lower energies
hinting at the previously mentioned red wing as manifestation of the joint effects close to the black hole leading to a
broadened and skewed shape.
We also explore the statistical difference between the narrow and the broad Gaussian used in addition to the 
broken power-law model. To establish the statistical significance of this narrow component we perform the very same
Monte Carlo simulations used for the broad component however keeping the fitted line width fixed to its best value.
As a result the difference in $\chi^{2}$ is significant to only $\sim$50\%.\\

\begin{figure}
\includegraphics[width=1.0\columnwidth]{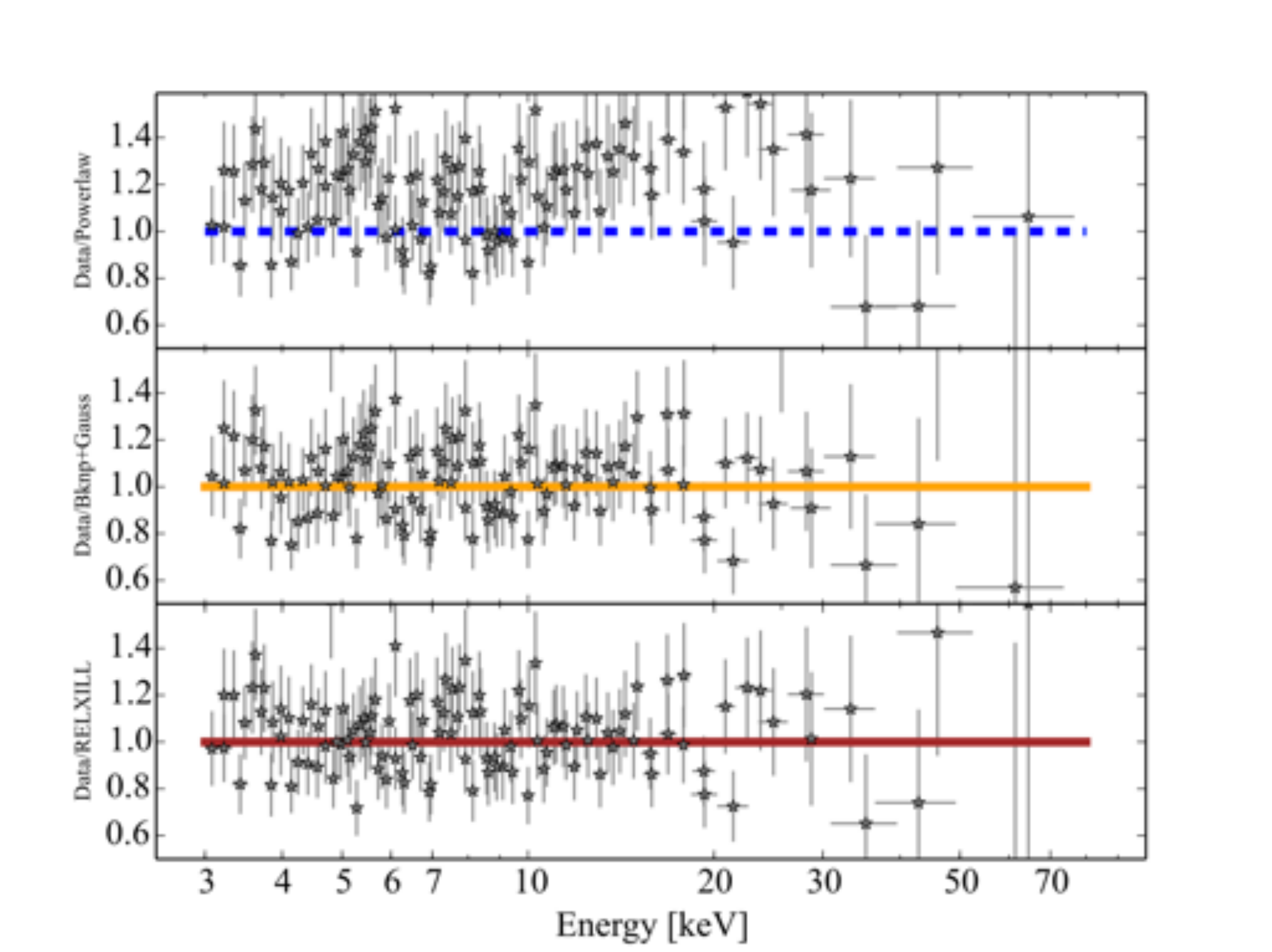}
\caption{Residuals (data/model) from top to bottom for the following models: {\texttt{powerlaw}} (dashed blue), {\texttt{bknp+Gauss}} (orange), {\texttt{RELXILL}} (brown), where
{\texttt{bknp}} stands for broken power law.
\label{fig:ratios}}
\end{figure}

\begin{figure}
\includegraphics[width=0.65\columnwidth,angle=-90]{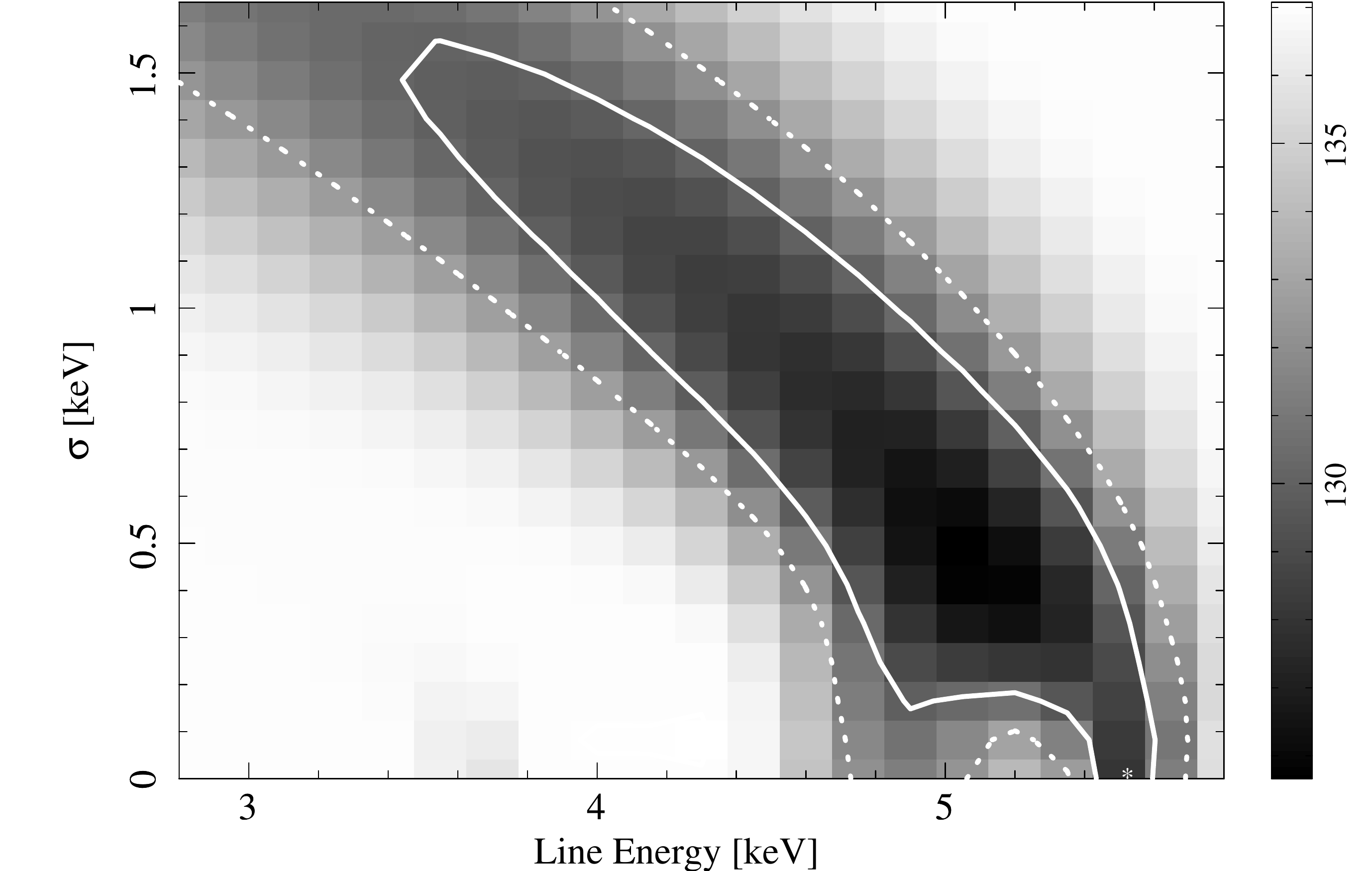}
\caption{Contours of the line energy (x-axis) and the line width (y-axis): solid and dashed lines delimit the 1$\sigma$ and 2$\sigma$ levels, respectively. The vertical colour
bar encodes the values of the $\chi^{2}$. The star at $\sim$ 5.5 keV represents the best fit value.
\label{fig:contours}}
\end{figure}

\subsection{Conclusion}

This research investigates Mrk~876's broadband spectrum as observed by {\em NuSTAR}. {\em NuSTAR}'s spectra
have been rebinned so that sharp spectral features are not smoothed out. This is important when exploring the spectral results
that show excesses with respect to a simple power-law model, which are characteristic for the reflection of the primary radiation
off the accretion disk. A turn over of the broadband continuum at 25.95 keV is interpreted as the Compton hump, while the
broad excess at low energies is coincident with the Fe K-shell emission line system. These excesses are less well fitted by
distant (from the SMBH) reflection models, which would  produce a narrow Fe line feature, rather than a broad feature.
In fact, the observed excess around the Fe line energy is best fitted by a broad ($\sigma$=1.4 keV) Gaussian component,
which is statistically significant at 99.71\% ($\sim$3$\sigma$) being the post-trial probability through Monte Carlo simulations.
The fit with a narrow component is significant to only $\sim$50\%.
This excludes the broad excess at low energies to be a statistical fluke.
A further complication for the distant reflection is the low value of the cut-off energy at E=11.59 keV, which is inconsistent
with theory and with current most sensitive measurements with {\em NuSTAR}.
The study of the low-energy excess shows that this system has a complex structure
displaying the red wing underpinning the possible combined effect of relativistic Doppler shift, relativistic beaming, and
gravitational redshift. The reflection model {\texttt{RELXILL}}
\citep[version v1.4.3;][]{garcia14, dauser14} represents the data rather well even though some parameters remain unconstrained
returning an upper limit for the black hole spin of $a$~$\leq$~0.85, while the inclination angle of the
accretion disk results in $i$=32.84\degree{}$^{+12.22}_{-8.99}$, which is in agreement within the errors with
a previous independent measurement ($i$=15.4$\degree{}^{+12.1}_{-6.8}$) by \cite{bian02}.
To confidently derive the parameters through the sophisticated reflection model, high-quality spectra are needed. Indeed, this
{\em NuSTAR} observation is of rather
low exposure ($\sim$30 kses) for such measurements, which prevents from obtaining well constrained fit parameters, even
though this model resembles well the best fit model (see Figure~\ref{fig:seds}). Especially the low-energy excess is well
fitted by this model when comparing the residuals (see Figure~\ref{fig:ratios}) thereby modeling the data due to the
red wing. For the spin measurement, Mrk~876's SMBH falls in an rather unique mass range
(2.4$\times$10$^{8}$~M$_\odot$~$\leq$~M$_{SMBH}$~$\leq$~1.3$\times$10$^{9}$~M$_\odot$), for which
possible moderately spinning SMBH are predicted \citep{reynolds13, vasudevan16} that are difficult to be measured. To
the best of our knowledge, the only other black holes mass exceeding the one in Mrk~876, for which a spin lower limit is
published, is H1821+643 \citep{reynolds14, reynolds21}.\\ It is worth pointing out that {\em NuSTAR} is able to detect the
spectral features associated to a rotating SMBH in Mrk~876 with a modest exposure of only $\sim$30 ksec compared
to much longer exposures for actual spin measurements for other sources \citep[e.g.][]{marinucci14}.\\
Furthermore, we report also the results of the analyses of 12 {\em Chandra} HETG observations of Mrk~876, which
hint at the soft excess at energies below $\sim$1.6 keV in agreement with previous analyses with {\em XMM}-Newton.
These observations also show, for the first time, significant variability at X-ray energies with an amplitude of 40\%. No
absorption in excess to the Galactic value is found thereby confirming previous findings.

\section*{Acknowledgements}
The author is grateful to the anonymous referee for their exhaustive and constructive criticism, which improved the
quality of the manuscript. 
The author is thankful to the {\em NuSTAR} team for performing the observations and for making data available.
This research has made use of the NuSTAR Data Analysis Software (NuSTARDAS) jointly developed by the ASI
Space Science Data Center (SSDC, Italy) and the California Institute of Technology (Caltech, USA). 
The TOPCAT tool \citep{taylor05} was used for this manuscript. This research has made use of data obtained from the
Chandra Data Archive and the Chandra Source Catalog, and software provided by the Chandra X-ray Center
(CXC) in the application packages CIAO and Sherpa.
The author acknowledges NASA grant 80NSSC21K0653.

\section*{Data Availability}
Observational data used in this paper are publicly available at NASA’s High Energy Astrophysics Science Archive
Research Center (HEASARC: https://heasarc.gsfc.nasa.gov/). Any additional information will be available upon
reasonable request.




\begin{thebibliography}{99}

\bibitem[Ananna et al.(2019)]{ananna19} Ananna, T.~T., Treister, E., Urry, C.~M., et al.\ 2019, \apj, 871, 240. doi:10.3847/1538-4357/aafb77

\bibitem[{{Arnaud}(1996)}]{arnaud96}
{Arnaud}, K.~A. 1996, in Astronomical Society of the Pacific Conference Series,
  Vol. 101, Astronomical Data Analysis Software and Systems V, ed. G.~H.
  {Jacoby} \& J.~{Barnes}, 17

\bibitem[Balokovi{\'c} et al.(2020)]{balokovic20} Balokovi{\'c}, M., Harrison, F.~A., Madejski, G., et al.\ 2020, \apj, 905, 41. doi:10.3847/1538-4357/abc342

\bibitem[Balokovi{\'c} et al.(2021)]{balokovic21} Balokovi{\'c}, M., Cabral, S.~E., Brenneman, L., et al.\ 2021, \apj, 916, 90. doi:10.3847/1538-4357/abff4d

\bibitem[Bambi(2021a)]{bambi21a} Bambi, C.\ 2021, arXiv:2106.04084

\bibitem[Bambi et al.(2021)]{bambi21b} Bambi, C., Brenneman, L.~W., Dauser, T., et al.\ 2021, \ssr, 217, 65. doi:10.1007/s11214-021-00841-8 

\bibitem[Baumgartner et al.(2013)]{baumgartner13} Baumgartner, W.~H., Tueller, J., Markwardt, C.~B., et al.\ 2013, \apjs, 207, 19. doi:10.1088/0067-0049/207/2/19
  
\bibitem[{{Bian} \& {Zhao}(2002)}]{bian02}
{Bian}, W., \& {Zhao}, Y. 2002, \aap, 395, 465

\bibitem[Bottacini et al.(2012)]{bottacini12} Bottacini, E., Ajello, M., \& Greiner, J.\ 2012, \apjs, 201, 34. doi:10.1088/0067-0049/201/2/34

\bibitem[Bottacini et al.(2015)]{bottacini15} Bottacini, E., Orlando, E., Greiner, J., et al.\ 2015, \apjl, 798, L14. doi:10.1088/2041-8205/798/1/L14

\bibitem[Boissay et al.(2016)]{boissay16} Boissay, R., Ricci, C., \& Paltani, S.\ 2016, \aap, 588, A70. doi:10.1051/0004-6361/201526982

\bibitem[Braito et al.(2007)]{braito07} Braito, V., Reeves, J.~N., Dewangan, G.~C., et al.\ 2007, \apj, 670, 978. doi:10.1086/521916

\bibitem[{{Cash}(1979)}]{cash79}
{Cash}, W. 1979, \apj, 228, 939

\bibitem[Comastri et al.(2015)]{comastri15} Comastri, A., Gilli, R., Marconi, A., et al.\ 2015, \aap, 574, L10. doi:10.1051/0004-6361/201425496

\bibitem[Dauser et al.(2014)]{dauser14} Dauser, T., Garcia, J., Parker, M.~L., et al.\ 2014, \mnras, 444, L100. doi:10.1093/mnrasl/slu125

\bibitem[{{Elvis} {et~al.}(1989){Elvis}, {Wilkes}, \& {Lockman}}]{elvis89}
{Elvis}, M., {Wilkes}, B.~J., \& {Lockman}, F.~J. 1989, \aj, 97, 777

\bibitem[{{Fabian} {et~al.}(2000){Fabian}, {Iwasawa}, {Reynolds}, \&
  {Young}}]{fabian00}
{Fabian}, A.~C., {Iwasawa}, K., {Reynolds}, C.~S., \& {Young}, A.~J. 2000,
  \pasp, 112, 1145

\bibitem[{{Fabian} {et~al.}(2012){Fabian}, {Zoghbi}, {Wilkins}, {Dwelly},
  {Uttley}, {Schartel}, {Miniutti}, {Gallo}, {Grupe}, {Komossa}, \&
  {Santos-Lle{\'o}}}]{fabian12}
{Fabian}, A.~C. {et~al.} 2012, \mnras, 419, 116

\bibitem[Fabian et al.(2015)]{fabian15} Fabian, A.~C., Lohfink, A., Kara, E., et al.\ 2015, \mnras, 451, 4375. doi:10.1093/mnras/stv1218

\bibitem[Fabian et al.(2017)]{fabian17} Fabian, A.~C., Lohfink, A., Belmont, R., et al.\ 2017, \mnras, 467, 2566. doi:10.1093/mnras/stx221

\bibitem[Faucher-Gigu{\`e}re(2020)]{faucher20} Faucher-Gigu{\`e}re, C.-A.\ 2020, \mnras, 493, 1614. doi:10.1093/mnras/staa302god

\bibitem[Fruscione et al.(2006)]{fruscione06} Fruscione, A., McDowell, J.~C., Allen, G.~E., et al.\ 2006, \procspie, 6270, 62701V. doi:10.1117/12.671760

\bibitem[Garc{\'\i}a \& Kallman(2010)]{garcia10} Garc{\'\i}a, J. \& Kallman, T.~R.\ 2010, \apj, 718, 695. doi:10.1088/0004-637X/718/2/695

\bibitem[Garc{\'\i}a et al.(2013)]{garcia13} Garc{\'\i}a, J., Dauser, T., Reynolds, C.~S., et al.\ 2013, \apj, 768, 146. doi:10.1088/0004-637X/768/2/146
  
 \bibitem[Garc{\'\i}a et al.(2014)]{garcia14}a, J., Dauser, T., Lohfink, A., et al.\ 2014, \apj, 782, 76. doi:10.1088/0004-637X/782/2/76
  
 \bibitem[Gehrels(1986)]{gehrels86} Gehrels, N.\ 1986, \apj, 303, 336. doi:10.1086/164079

\bibitem[{{George} \& {Fabian}(1991)}]{george91}
{George}, I.~M., \& {Fabian}, A.~C. 1991, \mnras, 249, 352

\bibitem[Gilli et al.(2007)]{gilli07} Gilli, R., Comastri, A., \& Hasinger, G.\ 2007, \aap, 463, 79. doi:10.1051/0004-6361:20066334

\bibitem[Harrison et al.(2013)]{harrison13} Harrison, F.~A., Craig, W.~W., Christensen, F.~E., et al.\ 2013, \apj, 770, 103. doi:10.1088/0004-637X/770/2/103

\bibitem[Hutchings \& Neff(1992)]{hutchings92} Hutchings, J.~B. \& Neff, S.~G.\ 1992, \aj, 104, 1. doi:10.1086/116216

\bibitem[Kallman \& Dorodnitsyn(2019)]{kallman19} Kallman, T. \& Dorodnitsyn, A.\ 2019, \apj, 884, 111. doi:10.3847/1538-4357/ab40aa

\bibitem[{{Kaspi} {et~al.}(2000){Kaspi}, {Smith}, {Netzer}, {Maoz}, {Jannuzi},
  \& {Giveon}}]{kaspi00}
{Kaspi}, S., {Smith}, P.~S., {Netzer}, H., {Maoz}, D., {Jannuzi}, B.~T., \&
  {Giveon}, U. 2000, \apj, 533, 631
  
  \bibitem[{{Kalberla} {et~al.}(2005){kalberla05}, {Burton}, {Hartmann}, {Arnal},
  {Bajaja}, {Morras}, \& {P{\"o}ppel}}]{kalberla05}
{Kalberla}, P.~M.~W., {Burton}, W.~B., {Hartmann}, D., {Arnal}, E.~M.,
  {Bajaja}, E., {Morras}, R., \& {P{\"o}ppel}, W.~G.~L. 2005, \aap, 440, 775
  
\bibitem[Kamraj et al.(2022)]{kamraj22} Kamraj, N., Brightman, M., Harrison, F.~A., et al.\ 2022, \apj, 927, 42. doi:10.3847/1538-4357/ac45f6  
 
\bibitem[Kang \& Wang(2022)]{kang22} Kang, J.-L. \& Wang, J.-X.\ 2022, \apj, 929, 141. doi:10.3847/1538-4357/ac5d49 

\bibitem[Kompaneets(1957)]{kompaneets57} Kompaneets, A.~S.\ 1957, Soviet Journal of Experimental and Theoretical Physics, 4, 730

\bibitem[Laha et al.(2014)]{laha14} Laha, S., Guainazzi, M., Dewangan, G.~C., et al.\ 2014, \mnras, 441, 2613. doi:10.1093/mnras/stu669

\bibitem[{{Lavaux} \& {Hudson}(2011)}]{lavaux11}
{Lavaux}, G., \& {Hudson}, M.~J. 2011, \mnras, 416, 2840

\bibitem[{{Lawson} \& {Turner}(1997)}]{lawson97}
{Lawson}, A.~J., \& {Turner}, M.~J.~L. 1997, \mnras, 288, 920

\bibitem[Lightman \& Zdziarski(1987)]{lightman87} Lightman, A.~P. \& Zdziarski, A.~A.\ 1987, \apj, 319, 643. doi:10.1086/165485

\bibitem[Marinucci et al.(2014)]{marinucci14} Marinucci, A., Matt, G., Kara, E., et al.\ 2014, \mnras, 440, 2347. doi:10.1093/mnras/stu404


\bibitem[Markowitz et al.(2006]{markowitz06} {Markowitz}, A., {Reeves}, J.~N., \& {Braito}, V. 2006, \apj, 646, 783

\bibitem[{{Matt} {et~al.}(1991){Matt}, {Perola}, \& {Piro}}]{matt91}
{Matt}, G., {Perola}, G.~C., \& {Piro}, L. 1991, \aap, 247, 25

\bibitem[Middei et al.(2019)]{middei19} Middei, R., Bianchi, S., Marinucci, A., et al.\ 2019, \aap, 630, A131. doi:10.1051/0004-6361/201935881

\bibitem[McKernan et al.(2007)]{McKernan07} McKernan, B., Yaqoob, T., \& Reynolds, C.~S.\ 2007, \mnras, 379, 1359. doi:10.1111/j.1365-2966.2007.11993.xNatalie Tom Plesa

\bibitem[Miller et al.(2008)]{miller08} Miller, L., Turner, T.~J., \& Reeves, J.~N.\ 2008, \aap, 483, 437. doi:10.1051/0004-6361:200809590

\bibitem[Miniutti et al.(2003)]{miniutti03} Miniutti, G., Fabian, A.~C., Goyder, R., et al.\ 2003, \mnras, 344, L22. doi:10.1046/j.1365-8711.2003.06988.x

\bibitem[Nandra et al.(2007)]{nandra07} Nandra, K., O'Neill, P.~M., George, I.~M., et al.\ 2007, \mnras, 382, 194. doi:10.1111/j.1365-2966.2007.12331.x

\bibitem[Nardini et al.(2016)]{nardini16} Nardini, E., Porquet, D., Reeves, J.~N., et al.\ 2016, \apj, 832, 45. doi:10.3847/0004-637X/832/1/45

\bibitem[{{Nayakshin} \& {Kazanas}(2001)}]{nayakshin01}
{Nayakshin}, S., \& {Kazanas}, D. 2001, \apjl, 553, L141

\bibitem[Oh et al.(2018)]{oh18} Oh, K., Koss, M., Markwardt, C.~B., et al.\ 2018, \apjs, 235, 4. doi:10.3847/1538-4365/aaa7fd

\bibitem[{{Piconcelli} {et~al.}(2005){Piconcelli}, {Jimenez-Bail{\'o}n},
  {Guainazzi}, {Schartel}, {Rodr{\'{\i}}guez-Pascual}, \&
  {Santos-Lle{\'o}}}]{piconcelli05}
{Piconcelli}, E., {Jimenez-Bail{\'o}n}, E., {Guainazzi}, M., {Schartel}, N.,
  {Rodr{\'{\i}}guez-Pascual}, P.~M., \& {Santos-Lle{\'o}}, M. 2005, \aap, 432,
  15

\bibitem[{{Porquet} {et~al.}(2004){Porquet}, {Reeves}, {O'Brien},\& {Brinkmann}}]{porquet04}{Porquet}, D., {Reeves}, J.~N., {O'Brien}, P., \& {Brinkmann}, W. 2004, \aap, 422, 85

\bibitem[{{Protassov} {et~al.}(2002){Protassov}, {van Dyk}, {Connors},
  {Kashyap}, \& {Siemiginowska}}]{protassov02}
{Protassov}, R., {van Dyk}, D.~A., {Connors}, A., {Kashyap}, V.~L., \&
  {Siemiginowska}, A. 2002, \apj, 571, 545
  
 \bibitem[Reeves et al.(2008)]{reeves08} Reeves, J., Done, C., Pounds, K., et al.\ 2008, \mnras, 385, L108. doi:10.1111/j.1745-3933.2008.00443.x

\bibitem[Reynolds(2013)]{reynolds13} Reynolds, C.~S.\ 2013, Classical and Quantum Gravity, 30, 244004. doi:10.1088/0264-9381/30/24/244004

\bibitem[Reynolds et al.(2014)]{reynolds14} Reynolds, C.~S., Lohfink, A.~M., Babul, A., et al.\ 2014, \apjl, 792, L41. doi:10.1088/2041-8205/792/2/L41

\bibitem[Reynolds(2016)]{reynolds16} Reynolds, C.~S.\ 2016, Astronomische Nachrichten, 337, 404. doi:10.1002/asna.201612321

\bibitem[Reynolds(2019)]{reynolds19} Reynolds, C.~S.\ 2019, Nature Astronomy, 3, 41. doi:10.1038/s41550-018-0665-z

\bibitem[Reynolds(2021)]{reynolds21} Reynolds, C.\ 2021, 43rd COSPAR Scientific Assembly. Held 28 January - 4 February, 43, 1412

\bibitem[Ricci et al.(2017)]{ricci17} Ricci, C., Trakhtenbrot, B., Koss, M.~J., et al.\ 2017, \apjs, 233, 17. doi:10.3847/1538-4365/aa96ad

\bibitem[Risaliti et al.(2013)]{risaliti13} Risaliti, G., Harrison, F.~A., Madsen, K.~K., et al.\ 2013, \nat, 494, 449. doi:10.1038/nature11938

\bibitem[{{Ross} \& {Fabian}(1993)}]{ross93}
{Ross}, R.~R., \& {Fabian}, A.~C. 1993, \mnras, 261, 74

\bibitem[{{Schmidt} \& {Green}(1983)}]{schmidt83}
{Schmidt}, M., \& {Green}, R.~F. 1983, \apj, 269, 352

\bibitem[Shull et al.(2011)]{shull11} Shull, J.~M., Stevans, M., Danforth, C., et al.\ 2011, \apj, 739, 105. doi:10.1088/0004-637X/739/2/105

\bibitem[Taylor(2005)]{taylor05} Taylor, M.~B.\ 2005, Astronomical Data Analysis Software and Systems XIV, 347, 29

\bibitem[{{Tombesi} {et~al.}(2010){Tombesi}, {Cappi}, {Reeves}, {Palumbo},
  {Yaqoob}, {Braito}, \& {Dadina}}]{tombesi10}
{Tombesi}, F., {Cappi}, M., {Reeves}, J.~N., {Palumbo}, G.~G.~C., {Yaqoob}, T.,
  {Braito}, V., \& {Dadina}, M. 2010, \aap, 521, A57
  
\bibitem[Tombesi et al.(2013)]{tombesi13} Tombesi, F., Cappi, M., Reeves, J.~N., et al.\ 2013, \mnras, 430, 1102. doi:10.1093/mnras/sts692

\bibitem[Tortosa et al.(2018)]{tortosa18} Tortosa, A., Bianchi, S., Marinucci, A., et al.\ 2018, \mnras, 473, 3104. doi:10.1093/mnras/stx2457

\bibitem[Treister et al.(2009)]{treister09} Treister, E., Urry, C.~M., \& Virani, S.\ 2009, \apj, 696, 110. doi:10.1088/0004-637X/696/1/110

\bibitem[Turner et al.(2000)]{turner00} Turner, T.~J., Perola, G.~C., Fiore, F., et al.\ 2000, \apj, 531, 245. doi:10.1086/308459

\bibitem[{{Turner} {et~al.}(2002){Turner}, {Mushotzky}, {Yaqoob}, {George},
  {Snowden}, {Netzer}, {Kraemer}, {Nandra}, \& {Chelouche}}]{turner02}
{Turner}, T.~J. {et~al.} 2002, \apjl, 574, L123

\bibitem[{{Turner} {et~al.}(2010){Turner}, {Miller}, {Reeves}, {Lobban},
  {Braito}, {Kraemer}, \& {Crenshaw}}]{turner10}
{Turner}, T.~J., {Miller}, L., {Reeves}, J.~N., {Lobban}, A., {Braito}, V.,
  {Kraemer}, S.~B., \& {Crenshaw}, D.~M. 2010, \apj, 712, 209

\bibitem[Vasudevan et al.(2013)]{vasudevan13} Vasudevan, R.~V., Brandt, W.~N., Mushotzky, R.~F., et al.\ 2013, \apj, 763, 111. doi:10.1088/0004-637X/763/2/111

\bibitem[Vasudevan et al.(2016)]{vasudevan16} Vasudevan, R.~V., Fabian, A.~C., Reynolds, C.~S., et al.\ 2016, \mnras, 458, 2012. doi:10.1093/mnras/stw363

\bibitem[Yee \& Green(1987)]{yee87} Yee, H.~K.~C. \& Green, R.~F.\ 1987, \aj, 94, 618. doi:10.1086/114495

\bibitem[Zdziarski et al.(1996)]{zdziarski96} Zdziarski, A.~A., Johnson, W.~N., \& Magdziarz, P.\ 1996, \mnras, 283, 193. doi:10.1093/mnras/283.1.193

\end{thebibliography}



\appendix


\bsp	
\label{lastpage}
\end{document}